# Analysis of Mission Opportunities to Sedna in 2029-2034

V.A. Zubko[a,b,], A.A. Sukhanov[a], K.S. Fedyaev[a], V.V. Koryanov[b], A.A. Belyaev[a,b]

[a]Department of Space Dynamics and Mathematical Information Processing, Space Research Institute of the Russian Academy of Sciences, 84/32 Profsoyuznaya Str, Moscow, 117997, Russian Federation
[b]Department of Dynamics and Flight Control of Rockets and Spacecrafts, Bauman Moscow State Technical University, 10 Gospitalnyi street, Moscow, 105005, Russian Federation

Abstract

The article focuses on trajectory design to the trans-Neptunian object (90377) Sedna for launch in 2029–2034. Sedna is currently moving to the perihelion at a distance of around 74 au from the Sun. The perihelion passage is estimated to be in 2073-74. That opens up of opportunities to study such a distant object. Known for its orbit and 10 thousand year period, Sedna is an exciting object for deep space exploration. The current research provides two possible scenarios of transfer to Sedna. A direct flight and flights including gravity assist manoeuvres are considered. The present study showed that a direct flight would be practically unrealistic due to the high total characteristic velocity and the time of flight value. Promising scenarios include gravity assist manoeuvres near Venus, Earth, Jupiter, Saturn and Neptune. The analysis of the close approach to asteroids during the flight to Sedna had been performed. Results of the research presented in this article show that the launch in 2029 provides the best transfer conditions in terms of minimum total characteristic velocity. The analysis shows that with a small additional impulses flybys of the large main-belt asteroids (16) Psyche for launch in 2034 and (20) Massalia for launch in 2029 are possible.

1. Introduction

A region of the Solar System located between 50 and 150 thousand au may be a source of material forming long-periodic comets. First, Strömgren (1947) and later Oort (1950) supposed the long-periodic comets that appeared during several centuries in the Sun's vicinity should have a source. In his work, E Strömgren described some of such comets. J.H. Oort generalised these data and proposed a concept to explain the origin of the phenomenon. According to his work (Oort, 1950), because in the previously mentioned region the solar gravity is weaker, the orbits of the bodies will be on the border of stability. Later, this region was named the Oort Cloud. Periodically, maybe due to the effect of the stellar passages near the Solar System, these objects' orbits become unstable. Thus under the solar gravity, they are pulled into the Solar System's interior and heating up (Oort, 1950; Duncan et al., 1987). The Oort Cloud objects probably are predominantly icy planetesimals, which can be formed to as comets, asteroids, and planetoids depending on their size (Kaib & Quinn, 2008).

For a long time, the Oort Cloud concept was only a hypothesis, but in 1992 astronomers J. X. X. Luu and D.C. Jewitt discovered the object (15760) Albion with perihelion about 40 au and aphelion about 43 au. This discovery confirmed existence of the Kuiper belt's region of the Solar System between about 30 and 60 au. The object (15874) 1996 TL$_{66}$ with aphelion of more than 80 au was discovered in 1996. The giant planets' gravity influence might modify the 1996 TL$_{66}$ orbit. The discovery revealed the necessity of classifying a distinct region of space, located between the outer Kuiper Belt region and the Oort Cloud's inner part. This region was named the scattered disk because the objects' orbits can be modified due to the

---

influence of the gravity fields of the giant planets and scattered in space as they were. Most of the known bodies of the scattered disk are in an orbital resonance with Neptune. A further step was the object 2003 VB$_{12}$ discovery by astronomers M. Brown, C. Trujillo, and D. Rabinowitz at about 80 au from the Sun (Brown et al., 2004). Later this object has been named as (90377) Sedna. From 2004 to 2006, a series of such large objects as Haumea, Makemake, Erida, Kwawar, Ork (about 1-2 thousand km in size) followed (Jewitt & Luu, 1995), as well as of the other currently known relatively small objects in the Kuiper Belt and the scattered disk were discovered. The discovery of so many large objects led to the logical withdrawal of Pluto from the planet category (Pluto and those large objects are now called dwarf planets) and, at the same time, to the establishment of a new group of bodies named trans-Neptunian.

The official organisation for receipt and distribution of positional measurements of minor planets, comets, and outer irregular natural satellites of the major planets, the Minor Planet Center[1], considers such objects as (148209) 2000 CR$_{105}$ and (90377) Sedna as objects of the scattered disk. However, the first object was discovered slightly earlier from the Sun than Sedna, although it has a relatively small diameter (about 285 km compared to Sedna's diameter of about 1000 km) and a shorter distance from the Sun than Sedna, 393 au at aphelion. Discoverers (Brown et al., 2004) had classified Sedna as an inner object of the Oort Cloud (de la Fuente Marcos & de la Fuente Marcos, 2014; Saillenfest et al., 2019) due to the Sedna's aphelion radius which is about 1000 au. As of now, the object's belonging to the Oort Cloud is hypothetical and needs a specific study. The possible candidates for the Oort Cloud object may be the sednoid (541132) Leleākūhonua (discovered in 2015, previously known as 2015 TG$_{387}$) (Sheppard et al., 2019), and asteroid 2017 MB$_7$ (discovered in 2017), since their aphelion radii exceed 2000 au.

Sedna is distinguished among other objects mentioned above by its highly elongated orbit and a much greater aphelion radius. The surface composition of the object is also typical for the Kuiper Belt (Cuk, 2004). As estimated in (Trujillo et al., 2005; Barucci et al., 2005; Emery et al., 2007; Pál et al., 2012; Trujillo & Sheppard, 2014), the object has a layer of hydrocarbon sediment produced by irradiated methane. The existence of the hydrocarbon sediment may be a reason why the Sedna's surface is a bright red shade (Cuk, 2004; Sheppard, 2010). Several studies have suggested that there may be an ocean beneath the surface of Sedna, heated by the object's internal heat (Husmann et al., 2006). Sedna's genesis is the subject for debate. The discoverers (Brown et al., 2004) have supposed that Sedna was created in the Solar System at the early stage of its evolution, and its orbit was changed because of dynamic effects that followed the Sun's formation within a dense stellar cluster (Brown et al., 2004; Morbidelli & Levison, 2004; Kenyon & Bromley, 2004; Adams, 2010; Kaib et al., 2011; Brasser & Schwamb, 2015). According to other versions, Sedna's orbit was changed by a stellar encounter (Kaib & Quinn, 2008) (e.g., the passing Scholz's star about 70 thousand years ago (Mamajek et al., 2015) at a distance of 52 thousand au from the Sun), or Sedna was captured from a low-mass star or a brown dwarf in interstellar space (Morbidelli & Levison, 2004).

According to the orbital asteroid database [2,3] currently Sedna is at about 80 au distance from the Sun. The passage of its perihelion is expected to be in 2073-74. Since Sedna's orbital period is over 10 thousand years, a unique opportunity is currently open to launch a spacecraft to Sedna and study it at a close range.

Space missions to such a distant object as Sedna is better to perform using gravity assists of the planets, i.e. to use gravitational fields of the planets for increasing the spacecraft orbital energy and for a purposeful changing trajectory. The article (McGranaghan et al., 2011) presents results of studies of transfer trajectories to Sedna (as well as four other trans-Neptunian objects) in 2015-2047. That paper considers an Earth-Jupiter-Sedna transfer, i.e. gravity assist manoeuvre only at Jupiter [4] is assumed. The transfer optimization,

---

[1] The Minor Planet Center (MPC) official web page: https://minorplanetcenter.net (accessed January 21, 2021)

[2] Orbital elements were obtained using Jet Propulsion Laboratory's Horizons orbital database on-line ephemeris service, http://ssd.jpl.nasa.gov/ (accessed November 11, 2020)

[3] Asteroid Orbital Elements Database, https://asteroid.lowell.edu/main/astorb (accessed November 11, 2020)

[4] The authors of the article (McGranaghan et al., 2011) state that the Cassini-Huygens mission to Saturn used Venus and Jupiter gravity assists, namely the Earth-Venus-Venus-Jupiter-Saturn (EVVJ-Saturn) scheme. In fact, the Earth-Venus-Venus-Earth-Jupiter-Saturn scheme was used in this mission. The inclusion of a Venus-Jupiter arc in any flight scheme, no matter what manoeuvres precede it, is always significantly worse (in terms of the ΔV budget required for the flight) than the Earth-Jupiter arc. However, based on their statement, the authors correctly concluded that a direct flight to Jupiter after Earth-start results in a lower total ΔV than the EVVJ scheme, and on that basis, used a direct flight to Jupiter. (We note that such a

in terms of minimum ΔV budget, was made by the authors for the time of flight of 24.48 yrs; this minimum, depending on the launch date, lies within the range from 7.2 (in 2045) to 11 (in 2023) km/s. Similar research has been conducted by (Kreitzman et al., 2013). In the paper, the authors studied use of the solar electric propulsion (low thrust) to reach distant trans-Neptunian objects.

There are also other publications devoted to the analysis of missions to trans-Neptunian objects. The article (Zangari et al., 2019) analyses the ΔV value required for a flight to one hundred of the objects using the Jupiter gravity assist. The paper (Sanchez et al., 2014, 2019) considers the flight to Haumea system and some other dwarf planets beyond the Neptune orbit, with the possibility of breaking into its satellite orbits. In the work (Sukhanov & de Almeida Prado, 2019) analysed the use of tethered swing-by for transfer to Haumea system. In work (Aime et al., 2021) the authors analysed the use of the fusion drive for fast transfer to the trans-Neptunian bodies.

The proposed paper presents the results of the detailed analysis of the optimal trajectories to Sedna in 2029, 2031, 2034. Also, launches in 2033, 2036, and 2037 are briefly considered, however, those years are less favourable for the mission than 2029, 2031, 2034.

The ΔV budget here and below would be designated as $\Delta V_\Sigma$. This value includes the $\Delta V_0$ value necessary for transfer from a low Earth orbit (LEO) to the heliocentric trajectory and manoeuvres using the on-board propulsion system (here and below, we will call these manoeuvres as active manoeuvres or impulses) near the planets or in the deep space. Any trajectory correction manoeuvres during the transfer are not taken into account, because they depend on the navigation and the manoeuvres execution accuracy.

The present paper provides results of study of various mission scenarios that include gravity assist manoeuvres near some planets. A direct Earth-to-Sedna flight is analysed, as well as other flight scenarios, which use the Venus-Earth-Earth-Jupiter or Venus-Earth-$\Delta V_\alpha$-Earth-Jupiter manoeuvre where $\Delta V_\alpha$ is an impulse near aphelion. Such a manoeuvre allows reaching Jupiter with the value of $\Delta V_\Sigma$ just a little bigger than the value of $\Delta V_0$ required only for Earth-Venus flight from LEO. Furthermore, all of the considered scenarios of flight to Sedna (except the direct flight) include a gravity assist manoeuvre near Jupiter. Gravity assist manoeuvres near Saturn and Neptune are also analysed. Uranus is located far away from the transfer trajectory for the considered launch period, this is why performing gravity assist of this planet is impossible.

Optimal (in terms of the minimum $\Delta V_\Sigma$ value) trajectories were determined for all the analysed flight scenarios. Optimal transfers of this type have a long duration; therefore, the restrictions on the flight duration equal to 20, 25, 30, 40, and 50 yrs were also considered. As will be seen below, the best flight scheme depends on the launch date: Earth-Venus-Earth-Earth-Jupiter-Sedna for 2029 and 2031, Earth-Venus-Earth-Earth-Jupiter-Neptune-Sedna for the 2034. In most of these cases, additional $\Delta V_\alpha$ near Earth-Earth aphelion slightly reduces $\Delta V_\Sigma$. Moreover, with the time of flight of 25 yrs (i.e. approximately the same as accepted in (McGranaghan et al., 2011; Kreitzman et al., 2013)), $\Delta V_\Sigma$ is significantly less than in all cases of trajectories to Sedna considered in (McGranaghan et al., 2011; Kreitzman et al., 2013).

Several flight schemes considered in this paper need approach to Jupiter to a close distance during the gravity assist. For example, this occurs in the Earth-Venus-Earth-Earth-Jupiter-Sedna transfer, with mission launch in 2034 and the time of flight of 30 yrs. In this case, the spacecraft flies by Jupiter at the height of 4.2 thousand kilometres, and powerful radiation belts of Jupiter may be dangerous for the electronic components of the spacecraft (Podzolko et al., 2011). However, the spacecraft passes the radiation belts fast during the Jupiter gravity assist, and radiation hazards to the electronic components of the spacecraft might be small. The authors of the proposed paper considered it reasonable to evaluate the $\Delta V_\Sigma$ value for launch in 2034 and the time of flight of 30 yrs with restrictions on the minimum height above Jupiter equal to 600 thousand km. An accurate estimation of the radiation dose received by the spacecraft during the Jupiter flyby and its danger for electronic components of the spacecraft, with consideration of the possible protection against radiation, requires an additional study.

Some of the analysed optimal trajectories to Sedna also show the possibility of close approach to one or more main belt asteroids with relatively small additional impulse required for close approach to the asteroids.

---

flight was used in the Pioneer-10 and 11, Voyager-1 and 2, and New Horizon missions). As a result, the authors obtained highly excessive values of total ΔV compared to the minimum possible values are shown in the current paper.

2. Mathematical Methods

Patched conic approximation as a model of the spacecraft motion is used (Breakwell & Perko, 1966; Hintz, 2015; Prado, 2007): it is supposed that during the spacecraft flight through the planets' spheres of influence (SOI) the spacecraft motion is affected only by the gravity force of the central body and beyond of the SOI (i.e. on the heliocentric trajectory) is influenced only by the gravity force of the Sun. In other words, all perturbing effects on the spacecraft motion (such as the attraction of other planets and other massive bodies, oblateness of the planets, solar radiation pressure) are neglected. Besides, when the spacecraft moves in the planets' SOI, size of the SOIs is considered infinite (Araujo et al., 2008) (i.e. the planetocentric velocity of the spacecraft on the SOI boundary is assumed to be equal to its asymptotic velocity). On heliocentric trajectories, the planetary SOI are considered as point ones (which is admissible in the first approximation due to the small sizes of these spheres compared to heliocentric distances); respectively, the time of flight in the sphere of influence of a planet is neglected [5].

Therefore, the n-body problem reduces to a set of two-body problems; the solution of each of them is a Keplerian orbit (i.e., a conic section). The problem is to find these Keplerian orbits and to join them (patch them together) into a flight trajectory (Battin, 1999).

Determination of the arcs of the heliocentric trajectories between each pair of celestial bodies encountered during the flight is done by solving the Lambert problem[6], which includes attaining an orbit basing on two positions and the time of flight between them. There are numerous methods for solving this problem; the authors used the method suggested in (Sukhanov, 1989). The trajectory arcs are determined in this way. The arcs incoming to and outgoing from the planet of the gravity assist manoeuvre determine the incoming and outgoing asymptotic velocities, respectively. In order to join these arcs the spacecraft flight in the planet's SOI is determined in the following way:

During the planet passage, the incoming and outgoing hyperbolas determined by the asymptotic velocities of arrival to and departure from the planet, respectively, are connected at their pericenters so that the tangents to them coincide. The magnitude of the difference of the velocity vectors at the pericenter of these hyperbolas is equal to the impulse $\Delta V$ needed to connect the trajectory's arrival and departure arcs. If the planet's gravitational field is incapable of turning the asymptotic velocity vector from its incoming direction to the outgoing one, an additional impulse rotating the velocity vector to the necessary direction is required.

During the passage of a small celestial body with a mass assumed to be zero, the value of the impulse connecting the arrival and departure parts of the trajectory is equal to the magnitude of the difference between the velocity vectors incoming to and outgoing from the body.

The patched conic approximation significantly simplifies the analysis and optimization of the flight trajectories at the same time providing acceptable accuracy at the early stages of the mission design (Brooks & Hampshire, 1971; Prado, 2007). As results of (Prado, 2007; Bradley & Russell, 2014) and our own estimation show, the difference between the minimum $\Delta V$ values obtained using the patched conic approximation and the accurate n-body problem solution is small, although the dates of launch and flybys corresponding to the values may be slightly different.

Optimization of the trajectory is carried out in the following way: the intervals of the dates of the launch from the Earth and the passage of each celestial body engaged in the flight (launch and flyby windows), containing the optimal dates, are defined . The optimal trajectory providing the minimum value of $\Delta V_\Sigma$ is then selected from the whole set of trajectories, corresponding to the dates from these intervals. The advantage of such an approach is in the easily implementing any restrictions on the flight, such as the time of flight for each heliocentric trajectory arc or the whole flight, the spacecraft's height of the planet flybys, ensuring the spacecraft's radio visibility from the Earth during each celestial body flyby etc.

---

[5] A more complicated version of the method of patched conic approximation assumes the sizes of the planetary SOI to be finite (Li et al., 2018)

[6] Generally, the design of space missions may require flights between celestial bodies and other types of trajectories, for example, a flight to a given point in space or ensuring specified trajectory parameters at the end of the mission. The authors' software allows to solve such problem s, but in the case of a flight to Sedna, it is enough to use the flights between celestial bodies.

For each gravity assist the optimization program determines whether the manoeuvre is unpowered or powered (i.e. with an impulse at the pericentre of the flyby trajectory). (Note that in all cases where it is possible, the unpowered flyby is optimal; this conclusion is confirmed by the extensive practice of designing interplanetary missions, but the authors have not found any theoretical proof.) In some cases, the gravitational field of the planet cannot rotate the incoming asymptotic velocity vector into the outgoing one. In such cases an additional impulse to turn this vector is required. The tables contained in the article give the values of both impulses in the spheres of influence of the planets.

3. Gravity assist manoeuvres for the flight to Sedna

Missions beyond Jupiter are appropriate to be performed by using the powerful gravity field of this giant planet [7]. The Jupiter gravity assist enables a significant increase of the spacecraft orbital energy and directing the spacecraft to a desired point in space. All the missions beyond the Jovian region, such as Pioneer-10 and 11, Voyager 1 and 2, Cassini, New Horizons, used Jupiter gravity assist; the Ulysses mission also used this manoeuvre to turn the orbit plane by 80° and to fly over the Sun's poles. In all these cases, except Cassini, a direct flight to Jupiter was used. Depending on the launch year, the minimum value of $\Delta V_\Sigma$ in such flight is 6.3 to 6.5 km/s in such flight, the optimal time of flight is two years or more.

Venus and Earth gravity assists may be used in order to reach Jupiter allowing a significant reduction of $\Delta V_\Sigma$. The Earth-Venus-Earth-Jupiter (EVEJ) transfer can be used (i.e., application of manoeuvre VEGA = Venus-Earth Gravity Assist) to lower $\Delta V_\Sigma$. However, the advantage of use of this manoeuvre compared to a direct flight to Jupiter in term s of the required $\Delta V_\Sigma$ is relatively small (less than 1 km/s) because a large additional active manoeuvre is required during the flyby of Venus or Earth; at the same time the EVEJ flight is much longer than the direct flight to Jupiter. This is why the EVEJ scheme was not considered in our analysis.

Note that preliminary analysis of the schemes including gravity assists can be made using the Tisserand graphs (Strange & Longuski, 2002), which allow to determine scheme providing a minimum $\Delta V$ value. In particular, as shown by the analysis of the flight schemes to Jupiter, carried out in the works (Strange & Longuski, 2002; Hughes, 2016) the lowest $\Delta V$ cost is provided by the Venus-Earth-Earth gravity assist. Apart from consideration of the direct Earth-to-Sedna and Earth-Jupiter-Sedna flights, we used this scheme, namely Venus-Earth-Earth Gravity Assists (VEEGA) in order to reach Jupiter. This scenario makes it possible to fly to Jupiter with $\Delta V_\Sigma$ value just slightly higher than the $\Delta V_0$ value necessary to fly from LEO to Venus. The time of flight between two Earth flybys is 2 or 3 years. A comparatively low impulse $\Delta V_\alpha$ near the aphelion of the Earth-to-Earth arc, reducing the value of $\Delta V_\Sigma$ in most cases, may be performed. Adding such a manoeuvre to the flight scenario was studied by the authors as well. Direct flight Earth-to-Sedna also has been considered for comparison with flights using gravity assist manoeuvres.

The VEGA and VEEGA manoeuvres have two disadvantages. First, they are only possible with a launch period of flight to Venus of every 1.6 years. Second, the VEGA manoeuvre increases the time of flight by a year and a half and the VEEGA manoeuvre by another 2 or 3 years. The first of these disadvantages can be eliminated by using the following flight scheme instead of these manoeuvres:

- launching the spacecraft into an elliptic heliocentric orbit with a high aphelion and a period of approximately two years;
- performing an active manoeuvre of a few hundred m/s near the aphelion;
- return to Earth at another point of its orbit and perform a gravity assist of the Earth (so-called $\Delta$VEGA manoeuvre = $\Delta V$+Earth Gravity Assist).

---

[7] By now the only exception is the Ulysses mission which used the Jupiter gravity assist, but was not performed beyond Jupiter.

The ΔVEGA manoeuvre may be performed on any launch date and has a significantly shorter duration than the VEEGA manoeuvre. However, in order to reach Jupiter ΔVEGA requires a ΔV of about 1.5 km/s more than VEEGA. The ΔVEGA manoeuvre was applied by the Juno spacecraft (Matousek, 2007), launched in 2011 to Jupiter, and currently researching this planet from a satellite orbit. The authors have not considered this manoeuvre in their analysis of the trajectories to Sedna. (It is easy to see that the VEEGA manoeuvre with $\Delta V_\alpha$ near the Earth-Earth trajectory aphelion is a combination of the VEGA and ΔVEGA manoeuvres).

The second disadvantage of the VEEGA manoeuvre can be easily avoided, thanks to the spacecraft's high orbital energy provided by the Venus, Earth, and Jupiter gravity assists. Due to this high energy, the total duration can be substantially reduced, with $\Delta V_\Sigma$ staying considerably lower than required for a direct flight to Jupiter and for the same time of flight; this will be demonstrated in this article.

The following designations will be used below for the scenarios of flight to Sedna:

ESed = Earth-Sedna direct flight
EJSed = Earth-Jupiter-Sedna
EVEEJSed = Earth-Venus-Earth-Earth-Jupiter-Sedna
EVEEJSSed = Earth-Venus-Earth-Earth-Jupiter-Saturn-Sedna
EVEEJNSed = Earth-Venus-Earth-Earth-Jupiter-Neptune-Sedna
EVEΔVE = Earth-Venus-Earth-Earth manoeuvre with an impulse $\Delta V_\alpha$ near the aphelion of the Earth-Earth trajectory segment.

The scenario of VEGA and ΔVEGA manoeuvres is illustrated by Fig. 1.

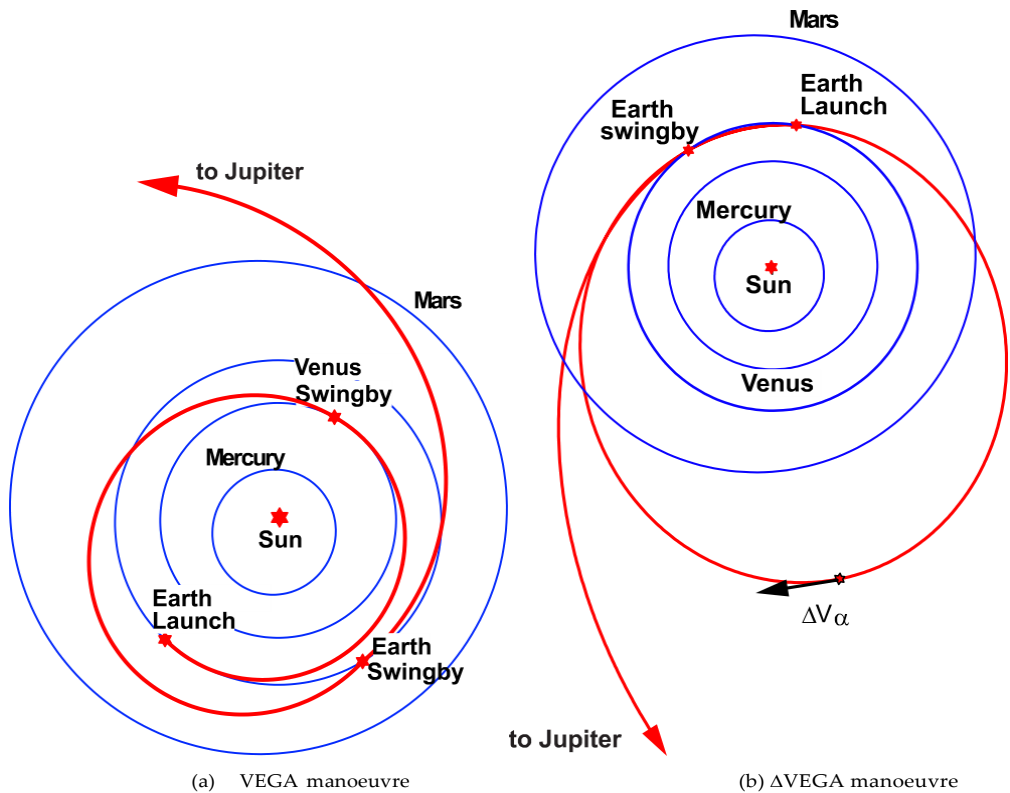

(a) VEGA manoeuvre  (b) ΔVEGA manoeuvre

Figure 1: The illustration of the VEGA and ΔVEGA manoeuvre.

4. Optimal flight trajectories to Sedna

This section presents the results of an optimal trajectory analysis for various flight schemes with constraints on the time of flight. Direct Earth-Sedna, Earth-Jupiter-Sedna flight, and flights with Venus, Earth, and Jupiter gravity assists are considered. The feasibility and expedience of gravity assist of the Saturn and Neptune are discussed as well. As seen from the tables below, there is a local minimum of $\Delta V_\Sigma$ as a function of the time of flight for EVEEJSed and EVE$\Delta$VEJSed schemes and launch in 2031 and 2034. If the mission begins in 2029, this minimum either does not exist or requires a time of flight 50-yrs or more.

In the subsequent analysis of flights with gravity assist manoeuvres, we will limit the time of flight by 50 yrs because assumed that a longer flight would be outside of the technical feasibility of such long-term space mission; the $\Delta V$ budget is restricted below by [8]

$$\Delta V_\Sigma \leq 8 km/s \qquad (1)$$

4.1. Direct flight to Sedna

This subsection presents the results of the direct flight to Sedna. The Earth-Sedna synodic period is about one year, this is why the launch to Sedna is possible every year. In this subsection we present the results of the mission analysis for the launch in 2029 with the time of flight limited by 50 yrs, The required $\Delta V_\Sigma$ versus time of flight to Sedna is shown in Fig. 2.

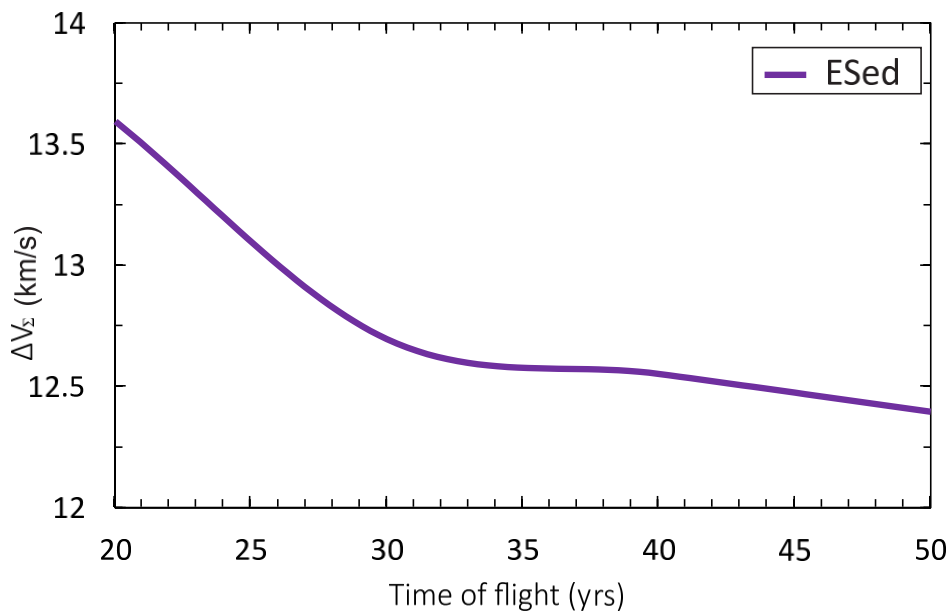

Figure 2: $\Delta V_\Sigma$ versus time of flight from 20 to 50 yrs for direct flight to Sedna for launch in 2029.

As seen in Fig. 2, for the time of flight considered in this paper, the minimum value of $\Delta V_\Sigma$ is more than 12 km/s.

For the direct flight within the entire range of considered launch dates and the time of flight not more than 50 years, the minimum $\Delta V_\Sigma$ is reached at 50 years of flight and exceeds 12 km/s. However, the time of flight can be significantly reduced by a relatively small increase of the $\Delta V_\Sigma$.

The direct flight to Sedna with extended time of flight interval was previously analysed in (Zubko et al., 2021). The minimum cost for the direct flight is achieved for the time of flight of 120 yrs or more, what the performed analysis shows (Fig. 3). The minimum value of $\Delta V$ is about 8.9 km/s that exceeds the taken constraint (1).

---

[8]The New Horizons mission's required 8.6 km/s for launch (Guo & Farquhar, 2008).

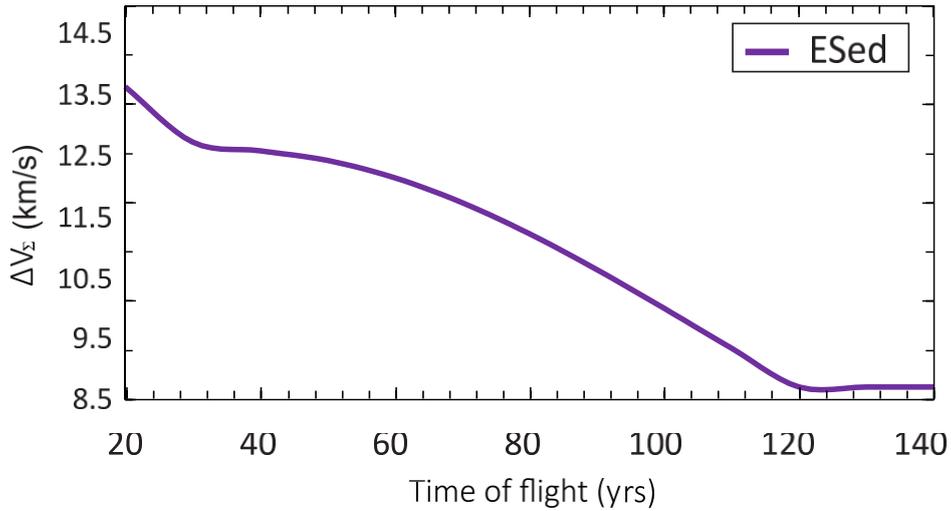

Figure 3: ΔV$_Σ$ versus time of flight from 20 to 120 yrs for direct flight to Sedna for launch in 2029.

4.2. Flight with Jupiter gravity assist

This subsection presents results of the analysis of the flight to Sedna with direct flight to Jupiter and gravity assist only of this planet. Since Jupiter's orbital period is 12 yrs, direct flights between Earth and Jupiter are possible every 13 months on average. In Fig. 4 total ΔV versus the time of flight for launch within considered time interval 2029-2034 are shown.

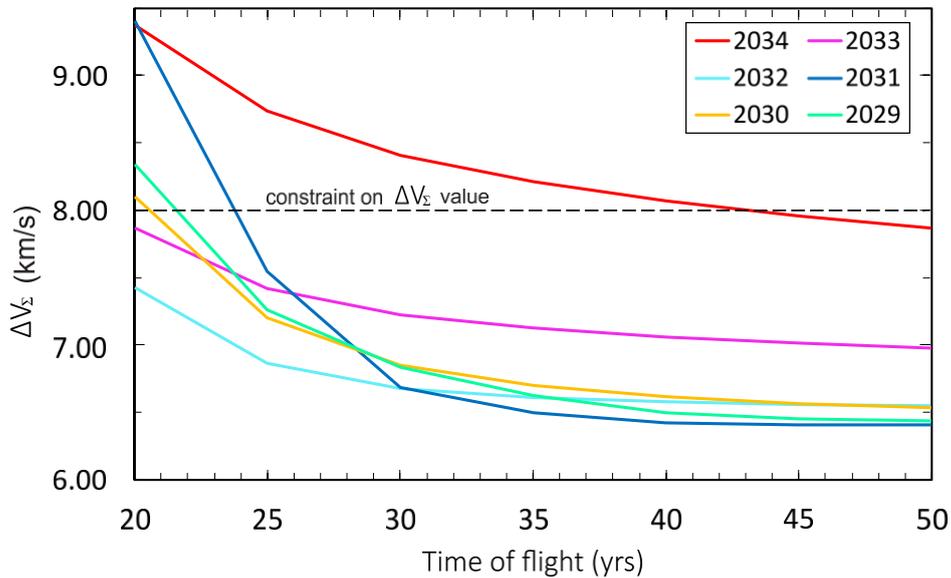

Figure 4: Total ΔV$_Σ$ versus the time of flight using the EJSed scheme in the launch date interval within 2029-2034.

As can be seen from Fig. 4 for the launches in 2029, 2030, 2031, and 2034, constraint (1) is satisfied for the time of flight more than 22, 21, 24, and 44 years, respectively.

As example in Table 1 the basic parameters of optimal flights within the period from 2029 to 2034 for time of flight of 30 years are shown.

Table 1 and Fig. 4 demonstrates that beginning from 2033, the value of ΔV$_Σ$ grows rapidly. That is explained by the position of Jupiter getting more and more unfavourable for the EJSed flights. Flight by this scenario becomes impossible in 2035, including flights longer than 30 yrs.

As seen in Table 1, the perijove height of the Jupiter flyby for launch in 2029 and 2030 (and to a lesser degree in 2031) is small. This may cause damage to the spacecraft electronic components by Jupiter's powerful radiation belts. This radiation hazard can be avoided by raising the height of the above Jupiter to a safer value; this raising will lead to increasing the $\Delta V_\Sigma$. The $\Delta V_\Sigma$ value versus the height for the launch in 2030 and the time of flight of 30 yrs is presented in Fig.5.

Table 1: Characteristics of the flights to Sedna using EJSed scheme.

| Optimal launch dates | Dates of the Jupiter flyby | Dates of arrival to Sedna | $\Delta V_\Sigma$, km/s | Height of the Jupiter flyby, $10^3$ km | Flyby velocity of Sedna, km/s |
|---|---|---|---|---|---|
| 2029 Feb 4 | 2033 Nov 3 | 2059 Feb 4 | 6.83 | 4 | 13.96 |
| 2030 Feb 6 | 2033 Sep 20 | 2060 Feb 6 | 6.84 | 3.6 | 13.34 |
| 2031 Feb 21 | 2032 Dec 11 | 2061 Feb 20 | 6.67 | 25 | 12.61 |
| 2032 Mar 28 | 2033 Nov 27 | 2062 Mar 25 | 6.68 | 388 | 12.23 |
| 2033 May 6 | 2034 Sep 21 | 2063 Apr 23 | 7.22 | 473 | 11.83 |
| 2034 June 18 | 2035 July 28 | 2064 May 30 | 8.41 | 2077 | 11.50 |

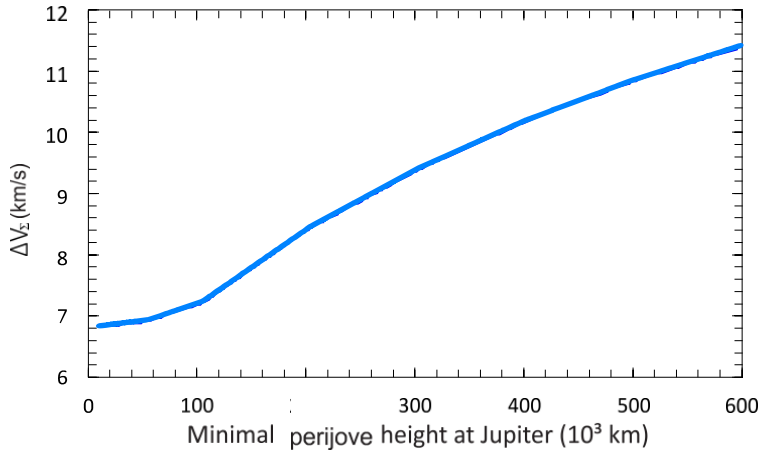

Figure 5: $\Delta V_\Sigma$ versus perijove height of the Jupiter flyby.

Comparison of Table 1 and Fig. 5 shows that in terms of the radiation hazard for the spacecraft near Jupiter the most favourable for the flight under the EJSed scheme within the considered period is the year 2032.

4.3. Launch in 2029

For the launch from Earth in 2029, the optimal flight is achieved by EVEEJSed (Zubko et al., 2021) or EVE$\Delta$VEJSed schemes; adding Saturn and Neptune gravity assists does not reduce $\Delta V_\Sigma$. Parameters of the flight trajectories are presented in Tables 2 and 3. The diagram on Fig. 6 shows $\Delta V_\Sigma$ versus time of flight.

As seen from the Table 2, an additional impulse for the turn of the velocity vector is not needed in all cases. In case of 20-yr flight duration the impulse near aphelion does not decrease $\Delta V_\Sigma$. The influence of such an impulse on $\Delta V_\Sigma$ grows with growing time of flight.

The $\Delta V_\Sigma$ value versus time of flight is illustrated by Fig. 6. The curve corresponding to the flight with $\Delta V_\alpha$ near the Earth-Earth trajectory aphelion (i.e. the EVE$\Delta$VEJSed scheme) is shown in blue, red curve corresponds to the flight without this impulse (i.e. the EVEEJSed scheme).

Table 2: Characteristics of the flights to Sedna using scheme EVEEJSed for launch in 2029.

| Time of flight, yrs | Optimal launch dates | $\Delta V_\Sigma$, km/s | $\Delta V$ of the velocity vector turn,* m/s | Height of the Jupiter flyby, $10^3$ km | Flyby velocity of Sedna, km/s |
|---|---|---|---|---|---|
| 20 | Nov 9 | 6.27 | 0 | 201 | 23.60 |
| 25 | Nov 1 | 5.05 | 0 | 484 | 17.50 |
| 30 | Oct 30 | 4.62 | 0 | 786 | 13.70 |
| 40 | Oct 28 | 4.32 | 0 | 1256 | 9.28 |
| 50 | Oct 28 | 4.20 | 0 | 1514 | 6.84 |

* No one of the gravity assist manoeuvres requre a velocity vector turn.

Table 3: Characteristics of the flights to Sedna using EVE$\Delta$VEJSed scheme for launch in 2029.

| Time of flight, yrs | Optimal launch dates | $\Delta V_\Sigma$, km/s | $\Delta V$ of the velocity vector turn,* m/s | $\Delta V_\alpha$ at aphelion, m/s | Height of the Jupiter flyby, $10^3$ km | Flyby velocity of Sedna, km/s |
|---|---|---|---|---|---|---|
| 20 | Nov 9 | 6.27 | 0 | 0 | 201 | 23.60 |
| 25 | Nov 3 | 5.02 | 0 | 121 | 483 | 17.50 |
| 30 | Oct 29 | 4.52 | 0 | 192 | 781 | 13.70 |
| 40 | Oct 27 | 4.16 | 0 | 313 | 1244 | 9.28 |
| 50 | Oct 27 | 4.00 | 0 | 303 | 1500 | 6.84 |

* No one of the gravity assist manoeuvres requre a velocity vector turn.

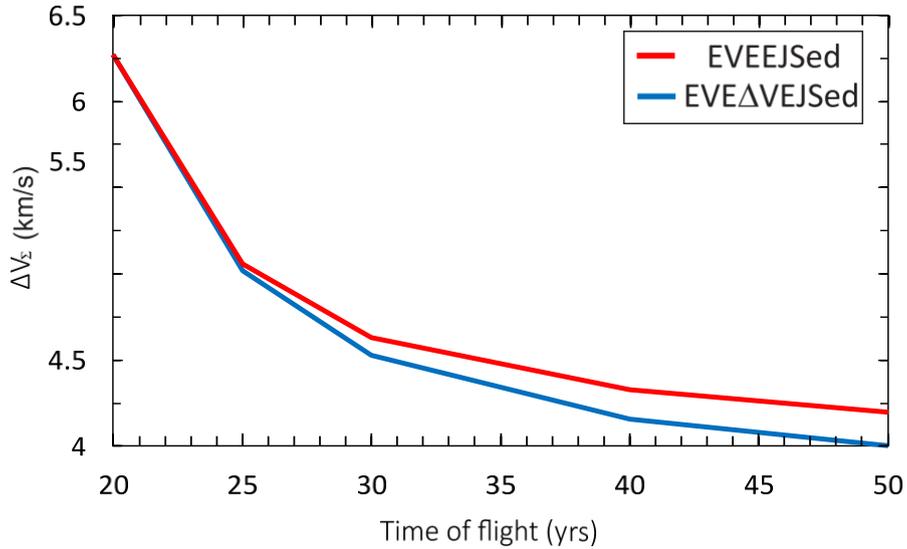

Figure 6: Total $\Delta V_\Sigma$ impulse versus time of flight (in years) for the launch in 2029.

Table 4 demonstrates more detailed results for the optimal flight using the EVE$\Delta$VEJSed scheme with the time of flight of 30 yrs. Corresponding transfer trajectory is shown in Fig. 7.

Table 4: Characteristics of the flight in 2029 using EVEΔVEJSed scheme with the time of flight of 30 yrs

| Celestial bodies | Dates of launch and flyby of celestial bodies | Relative velocities near Earth and of flyby of celestial bodies, km/s | ΔV of launch, at aphelion and of flyby of celestial bodies, km/s | ΔV of the velocity vector turn, m/s | Height of the initial orbit and flyby above celestial bodies, $10^3$ km |
|---|---|---|---|---|---|
| Earth | 2029 Oct 29 | 3.32 | 3.71 | – | 0.2 |
| Venus | 2030 Mar 27 | 5.53 | 0 | 0 | 5.9 |
| Earth | 2031 Feb 12 | 9.61 | 0 | 0 | 5.1 |
| ΔV$_\alpha$ | 2032 Apr 3 | – | 0.19 | – | – |
| Earth | 2033 May 26 | 10.69 | 0.7 | 0 | 0.5 |
| Jupiter | 2034 Sep 10 | 12.42 | 0 | 0 | 781 |
| Sedna | 2059 Oct 25 | 13.71 | – | – | 0 |

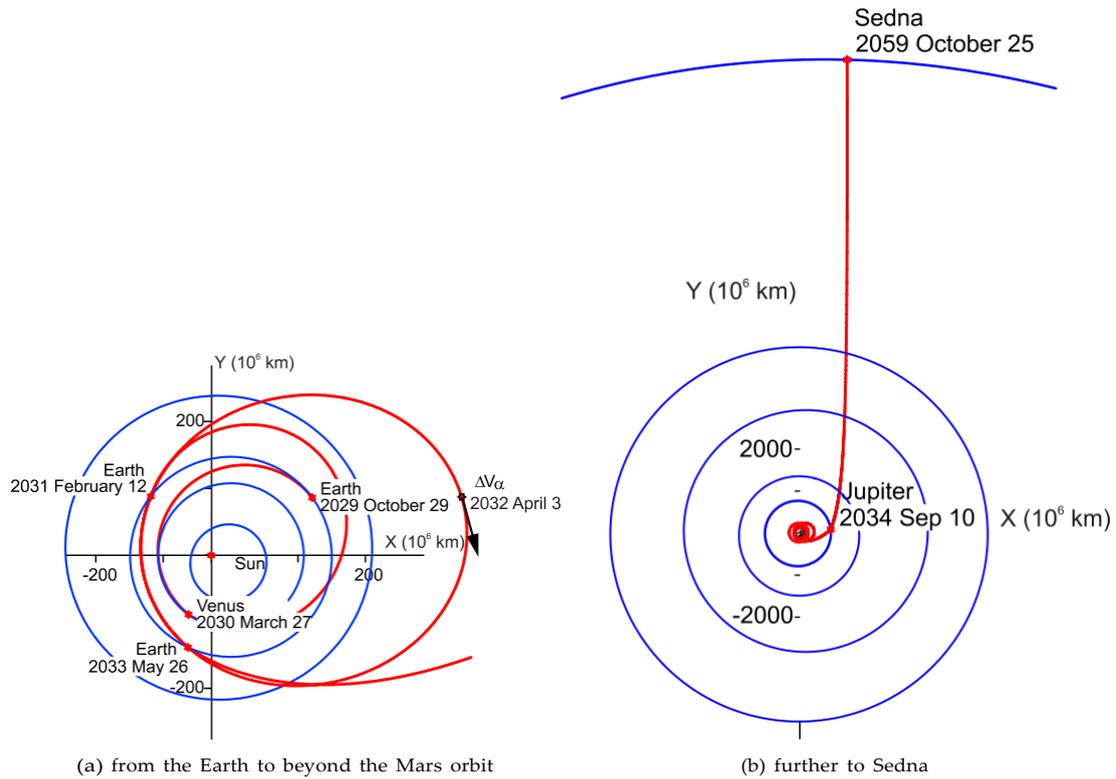

(a) from the Earth to beyond the Mars orbit  (b) further to Sedna

Figure 7: The flight trajectory to Sedna in 2029

4.4. Launch in 2031

For a launch in 2031, as well as in 2029, the optimal scheme is EVEEJSed or EVEΔVEJSed. Chracteristics of the optimal transfer trajectories are shown in Tables 5 and 6. The diagram of ΔV$_\Sigma$ versus the time of flight for these trajectories is shown in Fig. 8. As seen in Fig. 8, there are local minima of $\Delta V_\Sigma$ for the EVEEJSed scheme with the time of flight of about 36 yrs for the EVEEJSed scheme and with the time of flight of about 38 yrs for the EVE$\Delta$VEJSed scheme.

Table 5: Characteristics of the flights to Sedna using EVEEJSed scheme for launch in 2031.

| Time of flight, yrs | Optimal launch dates | $\Delta V_\Sigma$, km/s | $\Delta V$ of the velocity vector turn,* m/s | Height of Jupiter flyby, $10^3$ km | Flyby velocity of Sedna, km/s |
|---|---|---|---|---|---|
| 25 | July 28 | 8.18 | 153 | 3.6 | 34.42 |
| 30 | Aug 2 | 5.06 | 130 | 3.6 | 23.12 |
| 33 | Aug 3 | 4.36 | 127 | 3.9 | 19.10 |
| 35 | July 13 | 4.07 | 139 | 174 | 16.84 |
| 36.25** | July 6 | 3.97 | 82 | 292 | 15.68 |
| 40 | Dec 6 | 3.97 | 0 | 502 | 13.06 |
| 50 | June 3 | 4.07 | 0 | 577 | 8.90 |

*The turn of the velocity vector is performed at the first Earth flyby.
**The local minimum of $\Delta V_\Sigma$.

Table 6: Characteristics of the flights to Sedna using EVE$\Delta$VEJSed scheme for launch 2031.

| Time of flight, yrs | Optimal launch dates | $\Delta V_\Sigma$ km/s | $\Delta V$ of the velocity vector turn,* m/s | $\Delta V_\alpha$ at aphelion, m/s | Height of the Jupiter flyby, $10^3$km | Flyby velocity of Sedna, km/s |
|---|---|---|---|---|---|---|
| 25 | July 28 | 8.18 | 153 | 0 | 3.6 | 34.42 |
| 30 | Aug 2 | 5.06 | 130 | 0 | 3.6 | 23.12 |
| 33 | Aug 3 | 4.36 | 127 | 0 | 4.3 | 19.10 |
| 35 | July 10 | 4.06 | 114 | 25 | 172 | 16.84 |
| 38.51** | June 2 | 3.77 | 0 | 67 | 434 | 14.02 |
| 40 | June 2 | 3.78 | 0 | 74 | 501 | 13.07 |
| 50 | June 29 | 3.80 | 0 | 85 | 578 | 8.91 |

* The turn of the velocity vector is performed at the first Earth flyby.
** The local minimum of $\Delta V_\Sigma$.

As can be seen from Fig. 8 for both schemes constraint (1) is satisfied for the time of flight more than 26 years.

Details of the EVE$\Delta$VEJSed flight for the time of flight of 30 yrs are presented in Table 7. The corresponding trajectory is shown in Fig. 9.

As seen from Table 7, the spacecraft reaches Jupiter 10 yrs after the second Earth flyby and 12 yrs later than in 2029. The explanation is that Jupiter should complete a revolution around the Sun after 2029 to reach the position required for the gravity assist manoeuvre for the flight to Sedna. Fig. 9b shows that during the 10-yr Earth-Jupiter flight, the spacecraft almost reaches the Saturn's orbit; however, a Saturn gravity assist at this time is impossible, since Saturn is at a different point of its orbit.

Note that, as is seen in Table 7 and Fig. 9, for the launch in 2031, the optimal trajectory contains three-year interval between two Earth revisits, while in 2029 and, as will be seen below, in 2033 and 2034, this interval is two years.

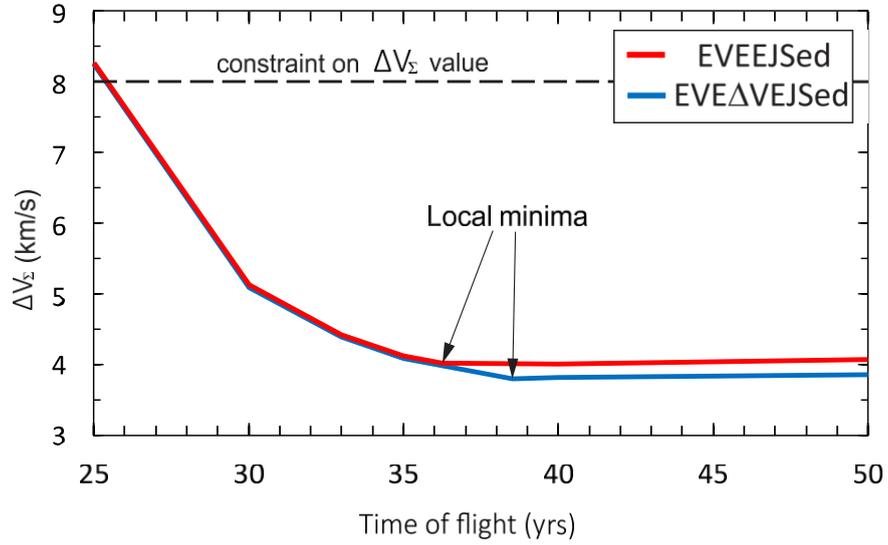

Figure 8: $\Delta V_\Sigma$ versus the time of flight (in years) for the launch in 2031.

Table 7: Characteristics of the flight in 2031 using EVE$\Delta$VEJSed scheme with the time of flight of 30 yrs

| Celestial bodies | Dates of launch and flyby of celestial bodies | Relative velocities near Earth and of flyby of celestial bodies, km/s | $\Delta V$ of launch, at aphelion and of flyby of celestial bodies, km/s | $\Delta V$ of the velocity vector turn, m/s | Height of the initial orbit and flyby above celestial bodies, $10^3$ km |
|---|---|---|---|---|---|
| Earth | 2031 Aug 2 | 4.24 | 4.01 | - | 0.2 |
| Venus | 2032 Jan 14 | 9.19 | 0 | 0 | 4.6 |
| Earth | 2032 Nov 25 | 11.50 | 0 | 140 | 0.3 |
| $\Delta V_\alpha$ | 2034 July 10 | - | 0 | - | - |
| Earth | 2036 Feb 23 | 11.49 | 0.09 | 0 | 0.7 |
| Jupiter | 2046 Feb 15 | 12.1 | 0.91 | 0 | 3.6 |
| Sedna | 2061 Aug 1 | 23.12 | - | - | 0 |

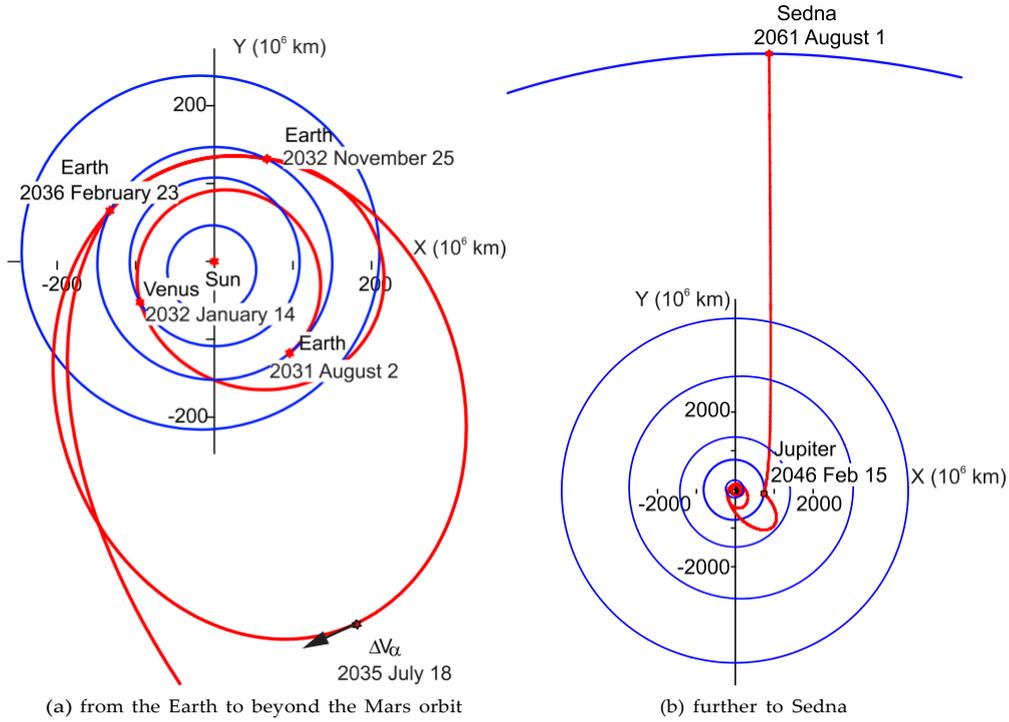

(a) from the Earth to beyond the Mars orbit    (b) further to Sedna

Figure 9: The flight trajectory to Sedna in 2031

4.5. Launch in 2033

For the launch in 2033, the $\Delta V_\Sigma$ value satisfies the accepted constraint (1) for the EVE$\Delta$VEJSed scheme only if the time of flight is not less than 33 yrs. Also, in the Jupiter-Sedna trajectory part the spacecraft approaches the Sun to a distance down to 22 million km. Besides, in all optimal flight scenarios in 2033 the spacecraft will pass above Jupiter at a very low height, namely no more than 3.6 thousand km (this value, equal to 5% of the radius of Jupiter, was chosen as the minimum possible), and may be exposed to radiation hazard. These issues make such a mission practically impossible.

However, both of these problems can be solved using the EVEEJSSed or EVE$\Delta$VEJSSed scheme with the Saturn gravity assist (see designations in Section 3). A significant disadvantage of these schemes is that a large acceleration impulse is required during the Saturn flyby, which grows rapidly with time of flight decreasing. As a result, condition (1) can be fulfilled only for the total time of flight more than 37 yrs.

Table 8 presents the parameters of the optimal flight using the EVE$\Delta$VEJSSed scheme with the duration of 40 yrs. The corresponding trajectory is shown in Fig. 10.

Neither of the gravity assist manoeuvres requires an additional impulse for the asymptotic velocity vector turn, as is shown in Table 8.

4.6. Launch in 2034

For the launch in 2034, the flight without the Neptune gravity assist (EVEEJSed or EVE$\Delta$VEJSed schemes) and one including such manoeuvre (EVEEJNSed or EVE$\Delta$VEJNSed schemes) is considered. The parameters of the flight trajectories are presented in Tables 9–12. The diagram of $\Delta V_\Sigma$ versus the time of flight for four schemes considered here is shown in Fig. 11.

As is seen in Tables 9–12 the turn of the asymptotic velocity vector is required only in the schemes including the Neptune flyby for the time of flight less than 27 yrs, and this turn is to be performed at the Neptune flyby. Tables 11 and 12 show that the impulse value necessary for this turn is large and leads to a significant increase in $\Delta V_\Sigma$.

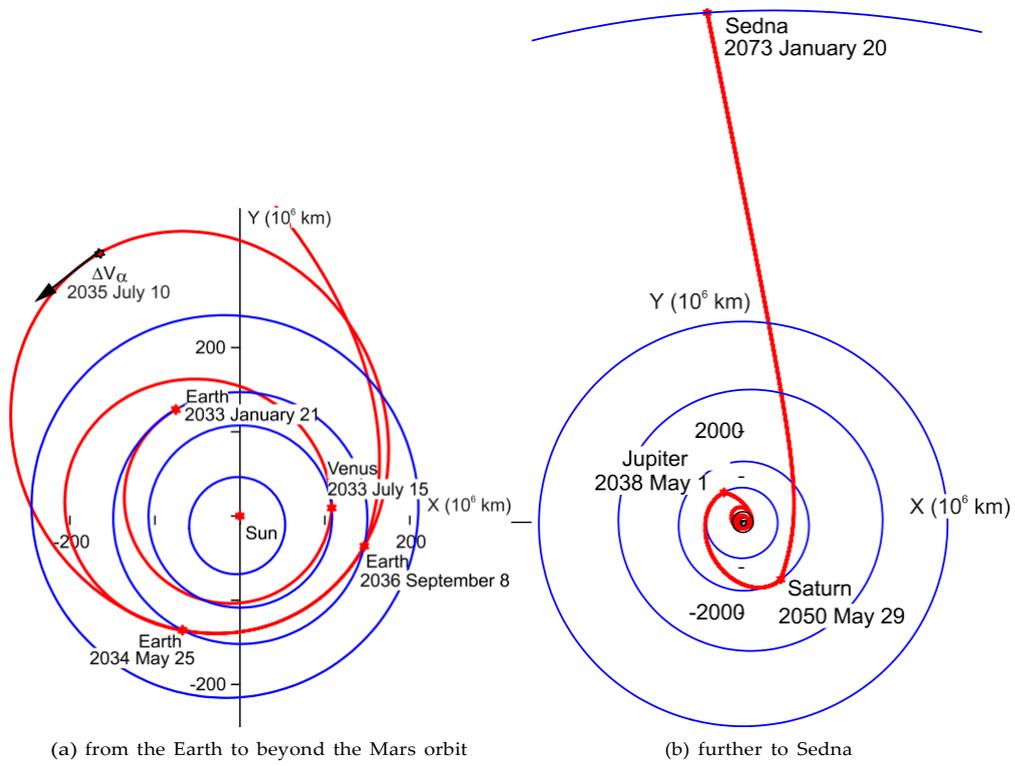

(a) from the Earth to beyond the Mars orbit

(b) further to Sedna

Figure 10: The flight trajectory to Sedna in 2033

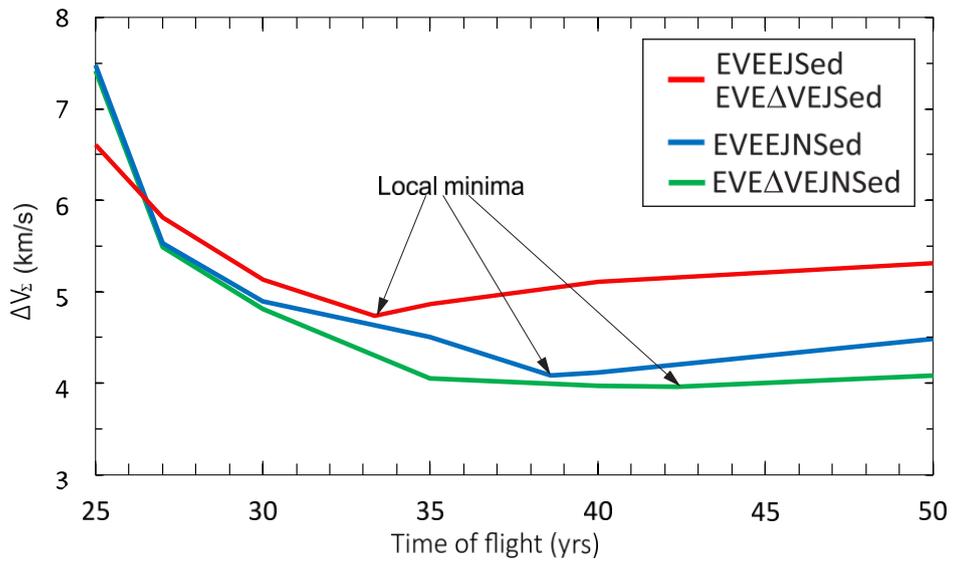

Figure 11: Total $\Delta V_\Sigma$ impulse versus time of flight (in years) for the launch in 2034 with and without a Neptune flyby.

Table 8: Characteristics of the flight in 2033 using EVEΔVEJSSed scheme with the time of flight of 40 yrs

| Celestial bodies | Dates of launch and flyby of celestial bodies | Relative velocities near Earth and of flyby of celestial bodies, km/s | ΔV of launch, at aphelion and of flyby of celestial bodies, km/s | ΔV of the velocity vector turn, m/s | Height of the initial orbit and flyby above celestial bodies, $10^3$ km |
|---|---|---|---|---|---|
| Earth | 2033 Jan 21 | 3.72 | 3.83 | - | 0.2 |
| Venus | 2033 July 15 | 6.14 | 0 | 0 | 8.5 |
| Earth | 2034 May 25 | 10.15 | 0 | 0 | 4.2 |
| $\Delta V_\alpha$ | 2035 July 18 | - | 0.11 | - | - |
| Earth | 2036 Sep 8 | 10.74 | 0 | 0 | 0.3 |
| Jupiter | 2038 May 1 | 9.05 | 0 | 0 | 739.9 |
| Saturn | 2050 May 29 | 5.83 | 2.71 | 0 | 18.7 |
| Sedna | 2073 Jan 20 | 16.55 | - | - | 0 |

Table 9: Characteristics of the flights to Sedna by the EVEEJSed scheme for launch in 2034.

| Time of flight, yrs | Optimal launch dates | $\Delta V_\Sigma$, km/s | ΔV of the velocity vector turn,* m/s | Height of the Jupiter flyby, $10^3$ km | Flyby velocity of Sedna, km/s |
|---|---|---|---|---|---|
| 25 | Aug 11 | 6.62 | 0 | 3.6 | 26.64 |
| 27 | Aug 8 | 5.82 | 0 | 3.6 | 23.08 |
| 30 | Aug 9 | 5.13 | 0 | 4.2 | 19.08 |
| 33.3** | Aug 8 | 4.73 | 0 | 4.8 | 15.87 |
| 40 | Aug 8 | 5.11 | 0 | 3.6 | 11.70 |
| 50 | Aug 7 | 5.31 | 0 | 3.6 | 8.15 |

* No one of the gravity assist manoeuvres requre a velocity vector turn.
** The local minimum of $\Delta V_\Sigma$.

The $\Delta V_\alpha$ value in the EVEΔVEJSed scheme is very small (see Table 10), this is why, as is seen in Tables 9 and 10 and Fig. 11, the $\Delta V_\Sigma$ values for the flights without and with impulse near aphelion are equal to each other. Tables 9–12 and Fig. 11 show that there are local minima of $\Delta V_\Sigma$ for all four flight schemes under consideration in this subsection. These minima are achieved at the times of flight of 33.3 yrs for the EVEEJSed and EVEΔVEJSed schemes, at 38.6 yrs. for the EVEEJNSed scheme and of 42.4 yrs. for the EVEΔVEJNSed scheme. As calculations show, the global minimum is reached at the time of flight over 100 yrs. for all four schemes.

As is seen in Fig. 11 that the blue curve corresponding to the EVEEJNSed transfer scheme has a kind of "hump" at the time of flight of about 35 yrs (see also Table 11). The authors were unable to find out the nature of this.

More detailed characteristics for the EVEΔVEJSed and EVEΔVEJNSed schemes with time of flight of 30 yrs are presented in Tables 13 and 14. Flight trajectories for these scenarios are shown in Fig. 12 and 13.

4.7. Remark on flights in 2031 and 2034

Some of the optimal trajectories by EVEEJSed and EVEΔVEJSed schemes in 2031 and 2034 include close Jupiter approach down to the accepted minimum perijove height of 3.6 thousand km (see Tables 5-7,

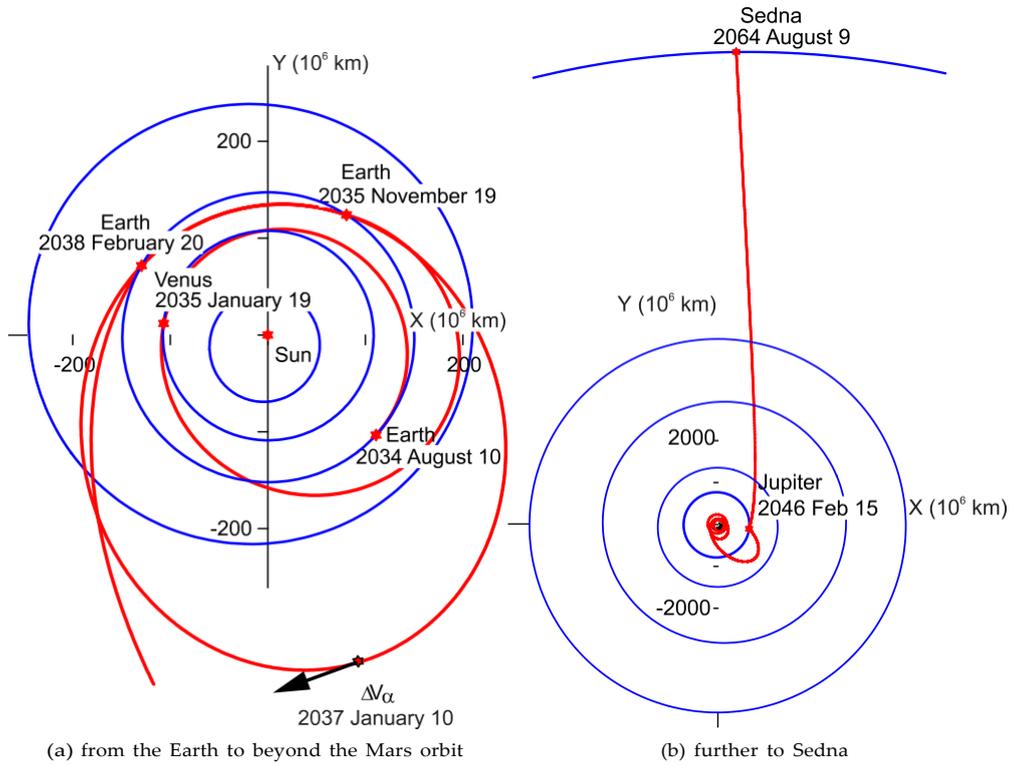

(a) from the Earth to beyond the Mars orbit    (b) further to Sedna

Figure 12: The flight trajectory using EVEΔVEJSed scheme to Sedna in 2034

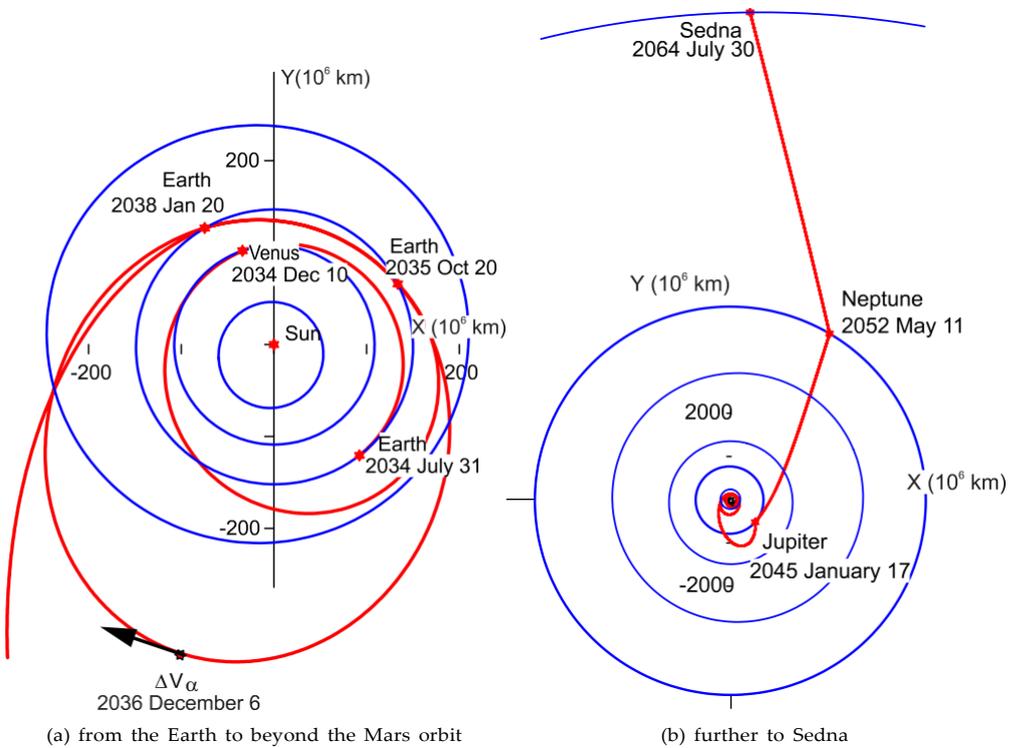

(a) from the Earth to beyond the Mars orbit    (b) further to Sedna

Figure 13: The flight trajectory using EVEΔVEJNSed scheme to Sedna in 2034

Table 10: Characteristics of the flights to Sedna by the EVEΔVEJSed scheme for launch in 2034.

| Time of flight, yrs | Optimal launch dates | $\Delta V_\Sigma$ km/s | $\Delta V$ of the velocity vector turn,* m/s | $\Delta V_\alpha$ at aphelion, m/s | Height of the Jupiter flyby, $10^3$ km | Flyby velocity of Sedna, km/s |
|---|---|---|---|---|---|---|
| 25 | Aug 11 | 6.62 | 0 | 8 | 3.6 | 26.64 |
| 27 | Aug 9 | 5.82 | 0 | 2 | 3.6 | 23.08 |
| 30 | Aug 10 | 5.13 | 0 | 1 | 3.6 | 19.07 |
| 33.3** | Aug 8 | 4.73 | 0 | 2 | 4.1 | 15.87 |
| 40 | Aug 7 | 5.11 | 0 | 4 | 3.6 | 11.70 |
| 50 | Aug 7 | 5.31 | 0 | 21 | 3.6 | 8.15 |

* No one of the gravity assist manoeuvres requre a velocity vector turn.
** The local minimum of $\Delta V_\Sigma$.

Table 11: Characteristics of the flights to Sedna by EVEEJNSed for launch in 2034.

| Time of flight, yrs | Optimal launch dates | $\Delta V_\Sigma$, km/s | $\Delta V$ of the velocity vector turn,* m/s | Height of the Jupiter flyby, $10^3$ km | Flyby velocity of Sedna, km/s |
|---|---|---|---|---|---|
| 25 | Aug 7 | 7.48 | 1222 | 3.7 | 27.40 |
| 27 | Aug 1 | 5.53 | 0 | 6.9 | 23.88 |
| 30 | July 30 | 4.89 | 0 | 32.2 | 19.82 |
| 35 | July 17 | 4.51 | 0 | 391 | 15.12 |
| 38.6** | Aug 5 | 4.08 | 0 | 673 | 12.74 |
| 40 | July 29 | 4.12 | 0 | 754 | 12.04 |
| 50 | July 3 | 4.48 | 0 | 1079 | 8.38 |

*A turn of the velocity vector is performed during the Neptune flyby only if the time of flight is less than 27 yrs.

** The local minimum of $\Delta V_\Sigma$.

9, 10); this is mostly typical for flights of relatively short duration. At a close height, the flyby of Jupiter may damage the spacecraft electronic components due to the passage through the powerful radiation belts of this planet. An increase of the perijove height needs a higher $\Delta V_\Sigma$ value (a similar problem already was considered for the Earth-Jupiter-Sedna (EJSed) scheme in subsection 4.2). The diagram of total $\Delta V_\Sigma$ versus the perijove height for the EVEEJSed scheme with launch in 2034 is presented in Fig. 14.

As seen in Fig. 14, an increase of the perijove height, for example, to 150 thousand km, will increase the cost of the characteristic velocity by about 500 m/s. However, we should also take into account the fact that in the EVEEJSed and EVE$\Delta$VEJSed schemes, the spacecraft approaches to Jupiter with a high asymptotic velocity (Tables 4, 7 and 8, 13 and 14) and flight through the Jupiter radiation belts takes a relatively short time, so the radiation hazard on the spacecraft may not affect on its functioning; this issue needs a special study.

4.8. Approach to asteroids on the transfer trajectory to Sedna

During the flight to Sedna, a close encounter with the asteroids of the main belt, which can increase the scientific output of the mission. Similar approaches to asteroids were carried out in the Galileo, Near, Cassini, and New Horizons missions (Porco et al., 2005; Veverka et al., 1997, 1999; Dunham et al., 2002; Olkin et al., 2006; Belton et al., 1996; Guo & Farquhar, 2008). An analysis of the possibility of close

Table 12: Characteristics of the flights to Sedna using scheme EVE$\Delta$VEJNSed for launch in 2034.

| Time of flight, yrs | Optimal launch dates | $\Delta V_\Sigma$ km/s | $\Delta V$ of the velocity vector turn,* m/s | $\Delta V_\alpha$ at aphelion, m/s | Height of the Jupiter flyby, $10^3$km | Flyby velocity of Sedna, km/s |
|---|---|---|---|---|---|---|
| 25 | Aug 11 | 7.41 | 1192 | 118 | 5.8 | 27.36 |
| 27 | Aug 3 | 5.49 | 0 | 79 | 4.1 | 23.88 |
| 30 | July 31 | 4.81 | 0 | 173 | 90 | 19.72 |
| 35 | July 27 | 4.05 | 0 | 387 | 407.2 | 15.05 |
| 40 | July 26 | 3.96 | 0 | 369 | 752 | 12.06 |
| 42.3** | July 22 | 3.96 | 0 | 378 | 873 | 10.96 |
| 50 | July 21 | 4.09 | 0 | 342 | 1062 | 8.37 |

* A turn of the velocity vector is performed during the Neptune flyby only if the time of flight is less than 27 yrs.

** The local minimum of $\Delta V_\Sigma$.

Table 13: Characteristics of the flight in 2034 using EVEΔVEJSed scheme with the time of flight of 30 yrs.

| Celestial bodies | Dates of launch and flyby of celestial bodies | Relative velocities near Earth and of flyby of celestial bodies, km/s | ΔV of launch, at aphelion and of flyby of celestial bodies, km/s | ΔV of the velocity vector turn, m/s | Height of the initial orbit and flyby above celestial bodies, $10^3$ km |
|---|---|---|---|---|---|
| Earth | 2034 Aug 10 | 3.87 | 3.88 | - | 0.2 |
| Venus | 2035 Jan 19 | 5.42 | 0 | 0 | 6.9 |
| Earth | 2035 Nov 19 | 9.24 | 0 | 0 | 4.4 |
| $\Delta V_\alpha$ | 2037 Jan 4 | - | 0 | - | - |
| Earth | 2038 Feb 20 | 9.23 | 0.84 | 0 | 0.3 |
| Jupiter | 2046 Feb 15 | 11.14 | 0.41 | 0 | 3.8 |
| Sedna | 2064 Aug 9 | 19.07 | - | - | 0 |

Table 14: Characteristics of the flight in 2034 using EVEΔVEJNSed scheme the time of flight of 30 yrs.

| Celestial bodies | Dates of launch and flyby of celestial bodies | Relative velocities near Earth and of flyby of celestial bodies, km/s | ΔV of launch, at aphelion and of flyby of celestial bodies, km/s | ΔV of the velocity vector turn, m/s | Height of the initial orbit and flyby above celestial bodies, $10^3$ km |
|---|---|---|---|---|---|
| Earth | 2034 July 31 | 3.65 | 3.81 | - | 0.2 |
| Venus | 2034 Dec 10 | 5.95 | 0 | 0 | 7.9 |
| Earth | 2035 Oct 20 | 8.79 | 0 | 0 | 4.5 |
| $\Delta V_\alpha$ | 2036 Dec 6 | - | 0.17 | - | - |
| Earth | 2038 Jan 21 | 9.78 | 0.43 | 0 | 0.3 |
| Jupiter | 2045 Jan 17 | 10.09 | 0.39 | 0 | 90.0 |
| Neptune | 2052 May 11 | 17.82 | 0 | 0 | 23.3 |
| Sedna | 2064 July 30 | 19.72 | - | - | 0 |

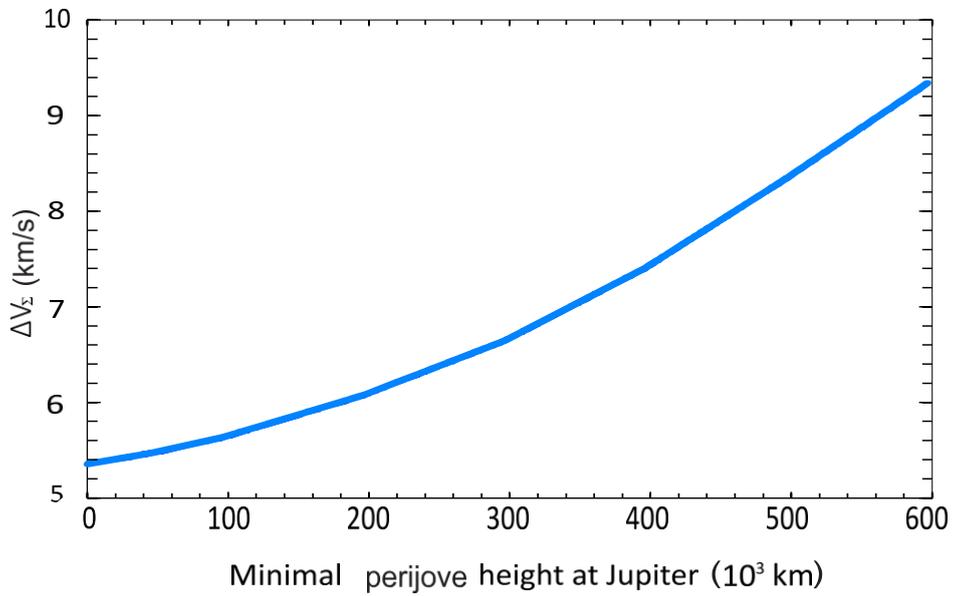

Figure 14: Total $\Delta V_\Sigma$ versus minimal perijove height at Jupiter during its flyby.

encounters with asteroids during the flight to Sedna was carried out for the launch in 2029, 2031 and 2034; only flight schemes EVEEJSed, as well as EVEEJNSed in 2034, were considered. The flight duration in all cases was taken equal to within 30 years. The approach is possible in two transfer parts: Earth-Earth and Earth-Jupiter. In the first case, the aphelion of the spacecraft trajectory reaches the main asteroid belt, and in the second case, the spacecraft trajectory crosses the entire main belt.

During the flight to Sedna a large number of asteroids have been found which the spacecraft can approach to with a relatively small additional ΔV necessary for such an approach (see Appendix A). Table 15 shows the main characteristics of the spacecraft's approach to one or two asteroids, at least one of which is quite large.

Table 11. Characteristics of close encounters with asteroids

| Year of launch | Asteroid | | Diameter, km | Date of flyby | Trajectory part | Flyby velocity, km/s | ΔV for one asteroid approach, m/s | ΔV for two asteroids approach, m/s |
|---|---|---|---|---|---|---|---|---|
| 2029 | 5080 | Oja | 12.1 | 8.12.2031 | E-E | 8.98 | 180 | 230 |
| | 20 | Massalia | 145.5 | 8.4.2032 | E-E | 5.67 | 220 | |
| 2031 | 1054 | Forsytia | 45.5 | 1.1.2034 | E-E | 9.23 | 20 | 250 |
| | 555 | Norma | 40.1 | 15.11.2036 | E-J | 12.92 | 170 | |
| 2034, EVEEJSed | 10520 | 1990 RS2 | 5.3* | 19.4.2037 | E-E | 4.99 | 245 | 120 |
| | 16 | Psyche | 253.2 | 9.10.2038 | E-J | 16.90 | 160 | |
| 2034, EVEEJNSed | 763 | Cupido | 17.3 | 12.4.2037 | E-E | 5.10 | 170 | 200 |
| | 153 | Hilda | 170.6 | 8.11.2038 | E-J | 15.19 | 30 | |

* The asteroid diameters were estimated using given absolute magnitudes of the asteroids and mean albedo value assumed to be equal to 0.1.

Given in two last columns of Table 15, the values of increase of $\Delta V_\Sigma$ required for close encounters with asteroids are an addition to the $\Delta V_\Sigma$ value for the flight without approaching asteroids (see Tables 2, 5, 9 and 11). During the analysis, these small additional ΔV were limited to

$$\Delta V \leq 250 m/s \qquad (2)$$

Let us consider the above results in more detail. As shown in Table 15, the total addition increase of $\Delta V_\Sigma$ necessary for the encounters with two asteroids can be either less than the sum of impulses for each asteroid separately (2029 and 2034 without the Neptune gravity assist), or bigger (2031). The reason lies in the mutual position of the asteroids. In 2031, the optimal trajectories, which include the encounter with only one of the two asteroids, differ significantly from each other, and their combination into one trajectory when approaching both asteroids requires a significant to increase of $\Delta V_\Sigma$. In 2029, on the contrary, the optimal trajectories with the flyby of each asteroid separately are relatively close to each other, and an increase of $\Delta V_\Sigma$ required to approach both asteroids is just slightly higher than such an increase to approach only one of the asteroids. When flying to Sedna in 2034 with the Neptune gravity assist, as follows from Table 15, the value of the impulse during the encounter with two asteroids is equal to the sum of the values necessary for the encounters with each of them separately; but this is just a coincidence.

A special case is the flight to Sedna in 2034, without flyby of Neptune, which includes close approaches to the asteroids. As shown in Table 15, the increase of $\Delta V_\Sigma$ for approaching both asteroids is significantly less than for approaching one of them. The reason is that in the cases of approaching one asteroid, the values of the increase of $\Delta V_\Sigma$ given in Table 15 were estimated using a one-impulse manoeuvre at the encounter point. However, for both asteroids (i.e. (16) Psyche and (10520) 1990 RS$_2$), a two-impulse manoeuvre is more optimal, in which for Psyche a second impulse is applied near the Earth-Earth trajectory part (where asteroid 1990 RS$_2$ is located during the approach to it), and for 1990 RS2 the best location for the application of the second impulse is probably in the area where Psyche is located during the encounter with her. Based on this, we should expect that two-impulse manoeuvres would also be advisable for approaching other asteroids, which would reduce the increase of $\Delta V_\Sigma$ given in Table 15. However, the study of optimal multi-impulse manoeuvres was beyond the scope of this analysis.

Note that the asteroid 1990 RS$_2$, which is a pair for Psyche, is small compared to other asteroids (about 5 km, see Table 15). The authors failed to find a larger asteroid, approach to which, together with Psyche, would require an increase of ΔV$_Σ$ less than the accepted limit of 250 m/s.

Consider how the approach to asteroids affects such an important parameter as the altitude of the Jupiter flyby. In 2029 and 2031, the inclusion of encounters with asteroids into the optimal trajectory almost does not change this altitude. But if for launch in 2029 this altitude is about 780 thousand km (see Table 2), then in 2031 it is equal to the accepted minimum permissible value of 3.6 thousand km (see Table 5) or slightly more (up to 3.9 thousand km). In 2034, for a flight without Neptune gravity assist, the inclusion of close approaches to asteroids raises the altitude of the Jupiter flyby from 4.2 thousand km (see Table 9) to 50 thousand km (approach only to Psyche and to both asteroids) or up to 80 thousand km (approach only to the 1990 RS$_2$ asteroid). In 2034, with the Neptune gravity assist, the inclusion of approaches to the asteroids lowers the altitude of the Jupiter flyby from 32 thousand km (see Table 11) to 3.6–6 thousand km.

## 5. Discussion

This section provides a comparison of approaches to solving the problem of flight to trans-Neptunian objects in the proposed article and in (McGranaghan et al., 2011; Kreitzman et al., 2013), and evaluates the advantages and disadvantages of our study.

In our study, we limited ourselves to a simplified approach: the patched conic approximation is used, the spacecraft mass is not estimated for specific launch vehicles and spacecraft propulsion systems. The $\Delta V_\Sigma$ value is used as a criterion for the feasibility and optimality of flights to Sedna. Also, the dose of radiation that the spacecraft may receive during its flyby of Jupiter is not estimated, however, the $\Delta V_\Sigma$ value versus the altitude of flight over Jupiter is obtained; this altitude is an important factor affecting the radiation dose. Such an approach is similar to the one used in (McGranaghan et al., 2011), while (Kreitzman et al., 2013) reviews the existing means of launching and manoeuvring in space, evaluates the initial and arrival mass of the spacecraft and analyses the radiation hazard during the Jupiter flyby taking into account different protection levels. Unlike papers (McGranaghan et al., 2011; Kreitzman et al., 2013), which consider flights to various trans-Neptunian objects using gravity assist only of Jupiter, in our study we focused on the mission to Sedna only, although at the same time we conducted a deeper analysis of this mission (within the framework of the simplified approach used). The analysis, the results of which are presented in this article, was carried out for four launch dates in 2029–2034 (these years seem to us the most realistic for organizing a flight to Sedna in the foreseeable future) and for various flight schemes, including both a direct flight to Sedna and from one to five gravitational manoeuvres, as well as a possible impulse near aphelion of the Earth-to-Earth part of the flight trajectory to Sedna. The values of $\Delta V_\Sigma$ given in the proposed article were obtained by simply summing the $\Delta V$ required for the transition from LEO to the heliocentric trajectory, and all $\Delta V_s$ performed by the spacecraft propulsion system. This summing is not quite correct, since these two types of manoeuvres are performed by different propulsion systems. For some of the flight options, the values of each of the manoeuvres are given in the article, which makes it possible to estimate $\Delta V_\Sigma$ more accurately for specific launch vehicles and spacecraft propulsion systems.

In (Kreitzman et al., 2013), flights to trans-Neptunian objects with both high and low thrust are considered. It is shown that the use of the low thrust makes it possible to somewhat increase the arrival mass of the spacecraft in comparison with the high thrust[9]. We limited ourselves to considering only the high thrust. However, as our analysis shows, a high-thrust flight according to the Earth-Venus-Earth-Earth-Jupiter-Sedna scheme can significantly increase the final mass of the spacecraft compared to a low-thrust flight according to the Earth-Jupiter-Sedna scheme for the same flight duration. Our analysis shows that a flight with high thrust under the EVEEJSed scheme leads on average to a doubling of the final mass of the spacecraft compared to a flight with low thrust under the EJSed scheme for the same time of flight, more than 25 years. At the same time, the EVEEJSed scheme considered in the proposed article leads to a significantly higher flyby velocity near Sedna than in the low thrust EJSed flight considered in (Kreitzman et al., 2013). This makes it practically impossible to orbit the spacecraft around Sedna as studied in (Kreitzman et al., 2013). Also the multiple gravity assists considered in our paper lead to a higher cost of the orbital correction manoeuvres than for the Earth-Sedna or Earth-Jupiter-Sedna flight. This cost was not estimated in the present paper.

The results obtained by us can serve as a basis for the development of a real project of the mission to Sedna, including the determination of the appropriate launch vehicle and propulsion system for the spacecraft. In contrast to (Kreitzman et al., 2013), which offers a limited set of these tools, their choice based on our results can be expanded.

6. Conclusions

As a result of the research possible flight schemes to the trans-Neptunian object (90377) Sedna were obtained and analyzed. Namely, the direct flight to Sedna, and the schemes including a series of gravity assists of the planets were considered.

As noted in subsection 4.1, the direct flight to Sedna is possible every year. For all optimal direct transfers to Sedna between 2029 and 2034 $\Delta V_\Sigma$ exceeds 12 km/s. The minimum for the direct flight of about 8.9 km/s is reached for the time of flight about 120 years (Fig. 3); i.e. even for such a long flight $\Delta V_\Sigma$ exceeds limit (1).

Among all the launch dates in the interval under consideration, 2029 is the most favourable for the EVEEJSed and EVE$\Delta$VEJSed flight schemes, since this is the only year when the $\Delta V_\Sigma$ value satisfies condition (1) when the time of flight is shorter than 25 yrs. Besides, if the time of flight does not exceed 30 years, then in 2029 $\Delta V_\Sigma$ is less than in any other launch year and any flight scheme.

In 2031 and 2034 there are local minima of $\Delta V_\Sigma$. In 2031 the minima are reached at the time of flight of 36.3 and 38.5 yrs. for the EVEEJSed and EVE$\Delta$VEJSed schemes respectively (see Fig. 8). In 2034 they are reached at 33.3 yrs. for the EVEEJSed and EVE$\Delta$VEJSed schemes and at 38.5 and 42.3 yrs. for the EVEEJNSed and EVE$\Delta$VEJNSed schemes respectively (see Fig. 11). Moreover, for the latter scheme, this minimum is 3.96 km/s which only slightly exceeds $\Delta V$. required for Earth-Venus flight Fig. 11 shows that when flying in 2034, the Neptune gravity assist leads to a lower value of $\Delta V_\Sigma$ compared to a flight without this manoeuvre with a flight duration of more than 27 years.

In 2033, a flight using the EVEEJSed and EVE$\Delta$VEJSed schemes is almost infeasible since the optimal trajectory approaches the Sun to a close distance (about 22 million km); besides the Jupiter gravity assist is performed at an height about 3.6 thousand km, which can lead to damage of the electronic components of the spacecraft by the powerful radiation belts of Jupiter. Both of these problems are resolved by adding Saturn gravity assist to the schemes mentioned above. It should be noted that without and with the Saturn gravity assist the constraint (1) is fulfilled with the time of flight not less than 33 and 37 years respectively.

As follows from Tables 5, 6, 9–12, some optimal trajectories to Sedna include the spacecraft approaching Jupiter at a short distance. To avoid this, the minimum height of Jupiter flyby can be limited from below by a certain safe value by means of some increase in the $\Delta V_\Sigma$ (see Fig. 14).

An analysis of the possibility of approaching asteroids during the flight to Sedna was also carried out; the results of the analysis are presented in this article. As an example, the approaches to large main belt asteroids (more than 140 km in diameter) (22) Massalia (start in 2029) (16) Psyche (2034 without Neptune gravity assist) and (153) Hilda (2034 with Neptune gravity assist) are possible. If the total value of the increase of $\Delta V_\Sigma$ required for a close encounter with the asteroids does not exceed 250 m/s (see condition (2)), then in each case, in addition to these asteroids, an encounter with one more asteroid of a smaller size is possible. For the launch in 2031, this limitation makes it possible to approach two asteroids of about 40 km in size (see Table 15).

When flying in 2034 without the Neptune gravity assist, an approach to one or both asteroids leads to a raise of the altitude of the Jupiter flyby from 4.2 thousand km to 50-80 thousand km, and when flying in the same year with the Neptune gravity assist and approaching one or both asteroids, this height lowers from 32 thousand km to 3.6-6 thousand km. At the launch in 2029 and 2031, the approach to the asteroids practically does not change the height of the Jupiter flyby.

Acknowledgements

Authors sincerely thank the participants of the Seminar on mechanics, control and informatics dedicated to trajectory design for promising interplanetary mission. The meeting led by Prof. R.R. Nazirov was held in the Space Research Institute (IKI) of the Russian Academy of Sciences. The seminar participants' comments helped to improve obtained results, which we subsequently included in this paper.

---

[9]The results of flight with high and low thrust given in Table 3 of (Kreitzman et al., 2013) are difficult to compare, as they are given for different input data, such as the launch vehicle used and the flight duration.

Appendix A. Candidates for close encounters among asteroids

Tables A.1 – A.8 show main belt asteroids, with one of which the spacecraft can approach to any small distance during its flight to Sedna. The search for such asteroids was carried out for the EVEEJSed flight scheme in 2029, 2031, and 2034, as well as for the EVEEJNSed scheme in 2034 (see the designations in Section 3). In this case, the approach to asteroids is possible on the Earth-to-Earth (Tables A.1, A.3, A.5, A.7) and the Earth-to-Jupiter (Tables A.2, A.4, A.6, A.8) trajectory parts. An additional impulse is needed for a close approach to an asteroid; approximate estimates of such impulses are given in the penultimate column of Tables A.1 – A.8. The values given in the column are significantly overestimated, since the estimation was carried out without optimizing the transfer trajectory including encounter with the asteroid. When evaluating the value of the additional impulse, a limitation of $\Delta V \lesssim 1$ km/s was imposed. On the optimal trajectory, including the approach to the asteroid, the additional impulse satisfies the accepted constraint (2) for practically all asteroids presented in the Tables. When compiling the Tables, only asteroids with a size of at least 5 km were considered.

Note that of all the asteroids listed in Table 15, only (10520) 1990 RS2 is presented in Table A.5. For the remaining seven asteroids in Table 15, the estimated additional impulse exceeds the accepted limit of 1 km/s. For example, for the asteroid (16) Psyche, this estimate is 5.1 km/s, while on the optimal trajectory this impulse is 160 m/s (see Table 15).

As shown in Tables 4, 8, 13, 14, the optimal flights to Sedna in 2029 and 2034 contain a two-year Earth-to-Earth flight, in which the aphelion of the orbit is about 2.3 AU; i.e. in these cases, the orbit covers the lower part of the asteroid belt. In 2031, the optimal flight has a three-year interval between two Earth passages (see Table 7). The aphelion distance of this segment of the trajectory is about 3.3 AU, i.e. the spacecraft almost reaches the upper boundary of the asteroid belt. This provides more opportunities for approaching asteroids during the Earth-Earth flight, which is confirmed by comparing the data in Table A.3 with Tables A.1, A.5 and A.7.

During the Earth-to-Earth flight the spacecraft encounters asteroids in the aphelion region, where the angle between the spacecraft velocity vector and the crossed asteroid orbits is relatively small, while the Earth-to-Jupiter trajectory crosses the asteroid orbits at a rather large angle. As a result, the flyby velocities of asteroids in the Earth-to-Earth segment are significantly lower than in the Earth-to-Jupiter segment, what is confirmed by the data given in the last column of Tables A.1-A.8.

Table A.1: Flight in 2029: Asteroids which can be encountered in the Earth-Earth trajectory part.

| Asteroid | | Diameter, km | Date of encounter | Flyby distance**, $10^6$ km | ΔV, km/s | Flyby velocity, km/s |
| --- | --- | --- | --- | --- | --- | --- |
| Number | Name | | | | | |
| 2442 | Corbett | 12.1* | 2032 May 16 | 4.050 | 0.79 | 6.97 |
| 6208 | Wakata | 6.7* | 2032 Apr 27 | 4.391 | 0.88 | 4.50 |
| 6940 | 1972 HL1 | 5.1* | 2032 Mar 1 | 4.322 | 0.88 | 4.92 |
| 8111 | Hoepli | 8.8* | 2032 Jan 26 | 5.480 | 0.68 | 9.01 |
| 11721 | 1998 HE100 | 7.3* | 2032 Mar 18 | 5.947 | 0.82 | 7.25 |
| 11790 | Goode | 8.4* | 2032 Mar 31 | 6.704 | 0.55 | 7.37 |
| 18376 | Quirk | 5.1* | 2032 Jul 31 | 4.303 | 0.82 | 10.46 |
| 22979 | 1999 VG25 | 6.4* | 2032 Jul 6 | 5.569 | 0.96 | 6.59 |
| 28498 | 2000 CL70 | 5.8* | 2032 Apr 11 | 6.426 | 0.85 | 7.54 |
| 29762 | Panasiewicz | 6.1* | 2032 Apr 15 | 2.398 | 0.28 | 5.63 |
| 59621 | 1999 JN72 | 5.1* | 2032 May 17 | 5.320 | 0.58 | 8.87 |
| 72473 | 2001 DD34 | 5.3* | 2032 Mar 7 | 2.748 | 0.34 | 7.44 |

\* The diameters of asteroids were estimated using given absolute magnitudes of the asteroids and mean albedo value assumed to be equal to 0.1.
\*\* This column of Tables A.1 – A.8 shows the closest distances to asteroids on the optimal trajectory of the spacecraft in the absence of any additional impulses to approach the asteroids.

Table A.2: Flight in 2029: Asteroids which can be encountered in the Earth-Jupiter trajectory part.

| Asteroid Number | Asteroid Name | Diameter, km | Date of encounter | Flyby distance**, $10^6$ km | $\Delta V$, km/s | Flyby velocity, km/s |
|---|---|---|---|---|---|---|
| 4061 | Martelli | 19.0 | 2033 Dec 11 | 3.912 | 0.62 | 14.24 |
| 4083 | Jody | 13.9* | 2033 Oct 22 | 4.069 | 0.73 | 21.25 |
| 6427 | 1995 FY | 8.0* | 2033 Sep 9 | 2.739 | 0.56 | 16.97 |
| 7208 | Ashurbanipal | 6.1* | 2033 Sep 22 | 1.070 | 0.20 | 16.14 |
| 7615 | 1996 TA11 | 6.7* | 2033 Sep3 | 4.461 | 0.98 | 19.91 |
| 8550 | Hesiodos | 14.6* | 2033 Dec 19 | 3.690 | 0.56 | 17.68 |
| 8867 | Tubbiolo | 7.6* | 2033 Dec 14 | 2.873 | 0.45 | 16.49 |
| 10176 | Gaiavettori | 5.1* | 2033 Aug 24 | 4.069 | 1.00 | 15.50 |
| 12474 | 1997 CZ19 | 6.4* | 2033 Sep 27 | 2.365 | 0.44 | 18.67 |
| 13638 | Fiorenza | 6.4* | 2033 Oct 7 | 2.320 | 0.42 | 16.08 |
| 15513 | Emmermann | 5.1* | 2033 Sep 22 | 1.201 | 0.23 | 16.96 |
| 22138 | Laynrichards | 5.5* | 2033 Nov 21 | 5.600 | 0.87 | 18.22 |
| 23598 | 1995 WL13 | 6.7* | 2033 Oct 22 | 2.676 | 0.47 | 18.61 |
| 26480 | 2000 AG198 | 6.7* | 2034 Jan 9 | 5.775 | 0.89 | 16.70 |
| 26608 | 2000 FZ33 | 5.8* | 2033 Oct 9 | 5.242 | 0.98 | 16.80 |
| 28604 | 2000 EB151 | 6.1* | 2033 Dec 6 | 6.008 | 0.95 | 13.78 |
| 29268 | 1993 FY22 | 7.0* | 2034 Jan 1 | 5.435 | 0.75 | 17.79 |
| 32215 | 2000 OG16 | 6.4* | 2033 Oct 26 | 5.925 | 1.00 | 17.10 |
| 36041 | 1999 QU | 6.4* | 2034 Feb 2 | 2.135 | 0.31 | 17.24 |
| 55922 | 1998 FL51 | 6.4* | 2034 Jan 4 | 6.313 | 0.96 | 18.58 |
| 59875 | 1999 RN114 | 5.3* | 2033 Dec 1 | 5.453 | 0.84 | 14.36 |
| 81214 | 2000 FL18 | 5.1* | 2033 Dec 1 | 2.097 | 0.31 | 17.79 |
| 231870 | 2000 SZ364 | 5.5* | 2034 Feb 23 | 5.848 | 0.83 | 15.24 |
| 244488 | 2002 TV19 | 5.3* | 2033 Dec 21 | 5.439 | 0.83 | 16.22 |

* The diameters of asteroids were estimated using given absolute magnitudes of the asteroids and mean albedo value assumed to be equal to 0.1.
** This column of Tables A.1 – A.8 shows the closest distances to asteroids on the optimal trajectory of the spacecraft in the absence of
any additional impulses to approach the asteroids.

Table A.3: Flight in 2031: Asteroids which can be encountered in the Earth-Earth trajectory part.

| Asteroid | | Diameter, km | Date of encounter | Flyby distance**, $10^6$ km | $\Delta V$, km/s | Flyby velocity, km/s |
|---|---|---|---|---|---|---|
| Number | Name | | | | | |
| 2103 | Laverna | 22.8 | 2034 Aug 6 | 6.582 | 0.61 | 6.24 |
| 3765 | Texereau | 11.6* | 2033 Dec 4 | 2.103 | 0.65 | 8.08 |
| 3768 | Monroe | 20.1* | 2035 May 25 | 2.384 | 0.78 | 10.12 |
| 5082 | Nihonsyoki | 10.6* | 2034 Jul 29 | 7.637 | 0.93 | 5.44 |
| 9473 | Ghent | 5.3* | 2035 Apr 1 | 3.630 | 0.21 | 8.94 |
| 12511 | Patil | 5.1* | 2033 Jun 17 | 3.801 | 0.69 | 16.01 |
| 15752 | Eluard | 10.1* | 2035 May 11 | 6.614 | 0.61 | 8.89 |
| 17038 | Wake | 6.4* | 2033 Aug 15 | 2.103 | 0.63 | 12.12 |
| 20403 | Attenborough | 8.4* | 2034 Feb 27 | 3.273 | 0.56 | 7.20 |
| 29074 | 5160 T-3 | 5.8* | 2035 Jun 20 | 10.306 | 0.84 | 14.90 |
| 35323 | 1997 CD26 | 5.1* | 2035 Feb 21 | 3.620 | 0.33 | 9.45 |
| 35425 | 1998 BY | 5.5* | 2035 May 30 | 5.216 | 0.35 | 10.97 |
| 36318 | 2000 LJ18 | 8.8* | 2035 Jul 31 | 5.579 | 0.68 | 7.66 |
| 51833 | 2001 OP47 | 7.0* | 2035 Mar 2 | 5.375 | 0.44 | 8.21 |
| 52552 | 1997 AD17 | 6.1* | 2034 Nov 6 | 6.635 | 0.76 | 6.22 |
| 63458 | 2001 OT6 | 8.0* | 2034 Jul 30 | 5.760 | 0.44 | 5.96 |
| 105131 | 2000 MD3 | 5.1* | 2034 Apr 17 | 7.262 | 0.97 | 6.90 |
| 111470 | 2001 YE8 | 5.3* | 2034 Sep 25 | 7.585 | 0.80 | 6.76 |
| 114649 | Jeanneacker | 5.3* | 2034 Jan 8 | 5.138 | 0.67 | 9.12 |
| 140855 | 2001 US219 | 5.8* | 2033 Dec 22 | 2.634 | 0.60 | 10.27 |
| 141355 | 2002 AK29 | 5.3* | 2035 Apr 5 | 4.205 | 0.67 | 7.83 |
| 160861 | 2001 FE186 | 5.1* | 2034 Dec 8 | 5.733 | 0.62 | 8.30 |
| 231467 | 2007 PZ19 | 5.3* | 2035 Apr 26 | 4.570 | 0.90 | 10.95 |

* The diameters of asteroids were estimated using given absolute magnitudes of the asteroids and mean albedo value assumed to be equal to 0.1.
** This column of Tables A.1 – A.8 shows the closest distances to asteroids on the optimal trajectory of the spacecraft in the absence of any additional impulses to approach the asteroids.

Table A.4: Flight in 2031: Asteroids which can be encountered in the Earth-Jupiter trajectory part.

| Asteroid | | Diameter, km | Date of encounter | Flyby distance**, $10^6$ km | $\Delta V$, km/s | Flyby velocity, km/s |
|---|---|---|---|---|---|---|
| Number | Name | | | | | |
| 4382 | Stravinsky | 13.9* | 2036 Aug 26 | 4.548 | 0.55 | 19.64 |
| 4479 | Charlieparker | 11.1* | 2036 Aug 5 | 7.380 | 0.93 | 16.22 |
| 10214 | 1997RT9 | 7.6* | 2036 Oct 23 | 9.601 | 0.89 | 16.40 |
| 10709 | Ottofranz | 8.4* | 2036 Sep 29 | 9.475 | 0.99 | 15.73 |
| 12569 | 1998VC29 | 12.7* | 2036 Oct 15 | 7.633 | 0.82 | 17.10 |
| 21080 | 1991RD18 | 5.8* | 2036 Jul 17 | 2.525 | 0.35 | 17.96 |
| 21966 | Hamadori | 5.5* | 2036 Jul 19 | 6.677 | 0.84 | 23.55 |
| 22904 | 1999TL19 | 5.1* | 2036 Aug 2 | 2.673 | 0.38 | 18.74 |
| 29688 | 1998XM92 | 6.7* | 2036 Jul 25 | 0.938 | 0.12 | 15.55 |
| 33721 | 1999LS34 | 6.4* | 2036 Oct 25 | 5.174 | 0.53 | 13.85 |
| 34045 | 2000OD34 | 5.8* | 2036 oct 25 | 3.902 | 0.31 | 14.33 |
| 38568 | 1999VE184 | 6.1* | 2036 Sep 21 | 4.652 | 0.47 | 18.00 |
| 44019 | 1997WO39 | 5.1* | 2036 Sep 15 | 4.545 | 0.50 | 18.04 |
| 44825 | 1999TS243 | 5.1* | 2036 Jul 15 | 8.519 | 0.97 | 20.86 |
| 58114 | 1981EL6 | 5.1* | 2036 Sep 4 | 10.056 | 1.00 | 17.04 |
| 70653 | 1999TP254 | 6.1* | 2036 Jul 31 | 5.541 | 0.64 | 17.49 |
| 70905 | 1999VF185 | 5.1* | 2036 Sep 16 | 3.938 | 0.46 | 14.26 |
| 77207 | 2001FE21 | 5.1* | 2036 Jul 25 | 3.369 | 0.59 | 19.25 |
| 91825 | 1999TM281 | 5.8* | 2036 Nov 25 | 10.932 | 0.99 | 15.66 |
| 92060 | 1999WY6 | 5.8* | 2036 Oct 19 | 10.041 | 0.94 | 17.47 |
| 92242 | 2000AO148 | 6.7* | 2036 Oct 6 | 7.061 | 0.74 | 17.69 |
| 164858 | 1999TX215 | 5.8* | 2037 Feb 4 | 10.794 | 0.85 | 17.16 |
| 249683 | 1999XD155 | 5.1* | 2036 Nov 22 | 4.126 | 0.37 | 17.45 |

* The diameters of asteroids were estimated using given absolute magnitudes of the asteroids and mean albedo value assumed to be equal to 0.1.
** This column of Tables A.1 – A.8 shows the closest distances to asteroids on the optimal trajectory of the spacecraft in the absence of any additional impulses to approach the asteroids.

Table A.5: Flight in 2034: Asteroids which can be encountered in the Earth-Earth trajectory part.

| Asteroid | | Diameter, km | Date of encounter | Flyby distance**, $10^6$ km | $\Delta V$, km/s | Flyby velocity, km/s |
|---|---|---|---|---|---|---|
| Number | Name | | | | | |
| 2195 | Tengstrom | 13.9* | 2036 Dec 23 | 6.599 | 0.85 | 5.37 |
| 2870 | Haupt | 11.1* | 2036 Nov 14 | 3.471 | 0.83 | 7.99 |
| 4788 | Simpson | 7.6* | 2036 Dec 16 | 4.108 | 0.58 | 5.77 |
| 7614 | Masatomi | 7.6* | 2036 Sep 26 | 2.960 | 0.97 | 6.51 |
| 9008 | Bohsternberk | 7.0* | 2037 Mar 5 | 4.362 | 0.83 | 4.79 |
| 10206 | 1997 PC2 | 5.8* | 2037 Apr 27 | 7.121 | 0.85 | 6.95 |
| 10520 | 1990 RS2 | 5.3* | 2037 Apr 27 | 12.838 | 0.79 | 3.53 |
| 19573 | Cummings | 5.8* | 2037 Jun 29 | 6.634 | 0.63 | 9.17 |
| 32154 | 2000 MH | 6.1* | 2036 Oct 11 | 7.615 | 0.96 | 11.54 |
| 136183 | 2003 UH208 | 5.8* | 2036 Dec 22 | 9.385 | 0.99 | 9.49 |

* The diameters of asteroids were estimated using given absolute magnitudes of the asteroids and mean albedo value assumed to be equal to 0.1.
** This column of Tables A.1 – A.8 shows the closest distances to asteroids on the optimal trajectory of the spacecraft in the absence of any additional impulses to approach the asteroids.

Table A.6: Flight in 2034: Asteroids which can be encountered in the Earth-Jupiter trajectory part.

| Asteroid | | Diameter, km | Date of encounter | Flyby distance**, $10^6$ km | Δ V, km/s | Flyby velocity, km/s |
|---|---|---|---|---|---|---|
| Number | Name | | | | | |
| 44 | Nysa | 70.6 | 2038 Aug 24 | 4.880 | 0.62 | 14.33 |
| 67 | Asia | 58.1 | 2038 May 7 | 2.861 | 0.44 | 20.68 |
| 189 | Phthia | 37.7 | 2038 Aug 7 | 3.557 | 0.51 | 17.28 |
| 2391 | Tomita | 13.3* | 2038 Sep 5 | 6.064 | 0.69 | 15.54 |
| 2700 | Baikonur | 16.7* | 2038 Oct 8 | 6.858 | 0.73 | 15.07 |
| 2861 | Lambrecht | 12.1* | 2038 Aug 24 | 3.273 | 0.37 | 16.48 |
| 4459 | Nusamaibashi | 7.3* | 2038 Aug 11 | 6.552 | 0.83 | 15.69 |
| 7683 | Wuwenjun | 6.7* | 2038 Aug 11 | 4.588 | 0.55 | 15.89 |
| 9609 | Ponomarevalya | 13.9* | 2038 Sep 18 | 7.263 | 0.91 | 17.27 |
| 10186 | Albeniz | 6.4* | 2038 Jun 28 | 4.151 | 0.75 | 20.58 |
| 11906 | 1992 AE1 | 8.8* | 2038 Oct 9 | 5.033 | 0.64 | 12.97 |
| 14468 | Ottostern | 5.1* | 2038 Jun 24 | 1.739 | 0.31 | 18.88 |
| 17040 | Almeida | 5.5* | 2038 Jul 26 | 3.949 | 0.51 | 16.52 |
| 17114 | 1999 JJ54 | 9.2* | 2038 Nov 6 | 2.878 | 0.27 | 13.38 |
| 18295 | Borispetrov | 10.6* | 2038 Aug 9 | 6.225 | 0.83 | 21.59 |
| 22640 | Shalilabaena | 5.1* | 2038 Jul 8 | 4.255 | 0.63 | 19.56 |
| 22883 | 1999 RC231 | 5.3* | 2038 Aug 26 | 6.046 | 0.70 | 22.61 |
| 27716 | Nobuyuki | 6.7* | 2038 Nov 11 | 9.294 | 0.91 | 11.31 |
| 27917 | Edoardo | 5.5* | 2038 Dec 3 | 6.692 | 0.63 | 16.82 |
| 42211 | 2001 DO49 | 6.4* | 2038 Sep 10 | 7.234 | 0.79 | 16.70 |
| 42778 | 1998 UC33 | 5.1* | 2038 Sep 24 | 10.135 | 0.92 | 18.52 |
| 45235 | 1999 XD228 | 6.1* | 2038 Oct 22 | 9.515 | 0.90 | 16.17 |
| 69374 | 1994 UH7 | 5.1* | 2038 Oct 11 | 7.845 | 0.88 | 15.68 |
| 72516 | 2001 DC81 | 6.1* | 2038 Aug 31 | 5.936 | 0.69 | 17.58 |
| 75231 | 1999 WZ4 | 5.3* | 2038 Aug 26 | 9.323 | 0.93 | 15.56 |
| 81484 | 2000 GF153 | 5.5* | 2038 Sep 29 | 2.250 | 0.24 | 16.93 |
| 83597 | 2001 SU263 | 7.6* | 2038 Sep 19 | 9.643 | 0.99 | 19.89 |
| 87242 | 2000 OU45 | 5.5* | 2038 Oct 22 | 5.451 | 0.54 | 18.01 |
| 88090 | 2000 WE48 | 7.3* | 2038 Nov 3 | 10.826 | 0.90 | 16.28 |
| 108872 | 2001 OE103 | 5.3* | 2038 Oct 27 | 6.668 | 0.69 | 18.35 |
| 140908 | 2001 VP53 | 5.1* | 2038 Nov 25 | 8.557 | 0.80 | 14.80 |
| 163586 | 2002 TQ189 | 5.5* | 2038 Oct 28 | 3.479 | 0.30 | 15.31 |

\* The diameters of asteroids were estimated using given absolute magnitudes of the
asteroids and mean albedo value assumed to be equal to 0.1.
\*\* This column of Tables A.1 – A.8 shows the closest distances to asteroids on
the optimal trajectory of the spacecraft in the absence of
any additional impulses to approach the asteroids.

Table A.7: Flight in 2034 with Neptune gravity assist: Asteroids which can be encountered in the Earth-Earth trajectory part.

| Asteroid | | Diameter, km | Date of encounter | Flyby distance**, $10^6$ km | Δ V, km/s | Flyby velocity, km/s |
|---|---|---|---|---|---|---|
| Number | Name | | | | | |
| 9174 | 1989 WC3 | 6.1* | 2036 Nov 15 | 4.127 | 0.88 | 4.37 |
| 16660 | 1993 US7 | 6.4* | 2037 Mar 14 | 4.358 | 0.56 | 7.30 |
| 40057 | 1998 KJ45 | 5.1* | 2036 Nov 27 | 2.159 | 0.27 | 8.46 |
| 45878 | 2000 WX29 | 5.1* | 2037 May 22 | 6.557 | 0.48 | 8.39 |

* The diameters of asteroids were estimated using given absolute magnitudes of the
asteroids and mean albedo value assumed to be equal to 0.1.
** This column of Tables A.1 – A.8 shows the closest distances to asteroids on
the optimal trajectory of the spacecraft in the absence of
any additional impulses to approach the asteroids.

Table A.8: Flight in 2034 with Neptune gravity assist: Asteroids which can be encountered in the Earth-Jupiter trajectory part.

| Asteroid | | Diameter, km | Date of encounter | Flyby distance**, $10^6$ km | $\Delta V$, km/s | Flyby velocity, km/s |
|---|---|---|---|---|---|---|
| Number | Name | | | | | |
| 153 | Hilda | 170.6 | 2038 Nov 4 | 9.220 | 0.95 | 15.57 |
| 962 | Aslog | 39.5 | 2038 Sep 21 | 3.237 | 0.35 | 15.96 |
| 1037 | Davidweilla | 8.0* | 2038 Jul 23 | 2.813 | 0.32 | 18.22 |
| 2527 | Gregory | 9.6* | 2038 Aug 9 | 4.290 | 0.46 | 18.29 |
| 3073 | Kursk | 8.0* | 2038 Jul 16 | 5.115 | 0.67 | 14.59 |
| 5036 | Tuttle | 16.7* | 2038 Aug 30 | 2.662 | 0.30 | 12.61 |
| 5682 | Beresford | 7.6* | 2038 Jul 19 | 8.001 | 0.95 | 20.77 |
| 5918 | 1991 NV3 | 11.1* | 2038 Sep 4 | 4.884 | 0.54 | 16.37 |
| 11464 | Moser | 5.3* | 2038 Dec 9 | 9.851 | 0.86 | 11.08 |
| 12262 | Nishio | 9.6* | 2038 Sep 29 | 2.308 | 0.21 | 15.00 |
| 13793 | Laubernasconi | 6.7* | 2038 Aug 16 | 5.552 | 0.62 | 18.67 |
| 17165 | 1999 LS27 | 8.4* | 2038 Sep 10 | 9.112 | 0.99 | 14.50 |
| 17276 | 2000 LU22 | 11.1* | 2039 Jan 1 | 12.466 | 0.81 | 15.20 |
| 18821 | Markhavel | 7.3* | 2038 Oct 5 | 3.601 | 0.32 | 15.35 |
| 22638 | Abdulla | 7.3* | 2038 Aug 11 | 1.888 | 0.20 | 17.81 |
| 25704 | Kendrick | 5.5* | 2038 Sep 3 | 3.605 | 0.36 | 16.13 |
| 27700 | 1982 SW3 | 5.5* | 2038 Jul 21 | 1.167 | 0.15 | 18.17 |
| 29939 | 1999 JS52 | 7.6* | 2038 Sep 17 | 8.339 | 0.91 | 15.16 |
| 30435 | 2000 LB29 | 9.6* | 2038 Oct 20 | 4.744 | 0.44 | 17.60 |
| 41359 | 2000 AG55 | 16.0* | 2039 Sep 5 | 9.401 | 0.80 | 10.01 |
| 41612 | 2000 SO124 | 6.7* | 2038 Nov 24 | 8.653 | 0.69 | 13.22 |
| 44388 | 1998 SK63 | 9.6* | 2038 Oct 3 | 7.872 | 0.87 | 13.06 |
| 46226 | 2001 HP2 | 9.2* | 2038 Sep 2 | 1.515 | 0.16 | 16.54 |
| 49478 | 1999 BY8 | 7.0* | 2038 Nov 5 | 0.903 | 0.07 | 13.89 |
| 57562 | 2001 TS48 | 6.7* | 2038 Aug 24 | 7.350 | 0.81 | 17.20 |
| 60136 | 1999 TV279 | 6.7* | 2038 Sep 1 | 6.362 | 0.71 | 18.06 |
| 67463 | 2000 QM204 | 5.3* | 2038 Aug 22 | 0.852 | 0.10 | 15.56 |
| 76666 | 2000 HV42 | 5.1* | 2038 Oct 31 | 10.428 | 0.95 | 12.08 |
| 83800 | 2001 TJ217 | 6.4* | 2038 Oct 17 | 9.775 | 0.91 | 14.04 |
| 98002 | 2000 QG199 | 8.0* | 2038 Oct 28 | 6.843 | 0.68 | 15.59 |

\* The diameters of asteroids were estimated using given absolute magnitudes of the asteroids and mean albedo value assumed to be equal to 0.1.
\*\* This column of Tables A.1 – A.8 shows the closest distances to asteroids on the optimal trajectory of the spacecraft in the absence of any additional impulses to approach the asteroids.